\def\imat {i}

\documentclass[prb,twocolumn,showpacs,aps]{revtex4}
\usepackage{graphicx}

\begin{document}

\title{Quantum Zeno stabilization in weak continuous measurement of two qubits}

\author{Rusko Ruskov,$^1$\footnote{On leave of absence from Institute
for Nuclear Research and Nuclear Energy, Sofia BG-1784, Bulgaria}
Alexander N. Korotkov,$^2$ and Ari Mizel$^1$}
\affiliation{$^1$Department of Physics and Materials Research
Institute, Penn State University, University Park, Pennsylvania
16802, U.S.A. \\
$^2$Department of Electrical Engineering, University of
California, Riverside, CA 92521-0204, U.S.A. }

\date{\today}


\begin{abstract}
We have studied quantum coherent oscillations of two qubits under
continuous measurement by a symmetrically coupled mesoscopic detector.
The analysis is based on a Bayesian formalism that is applicable to
individual quantum systems.
Measurement continuously collapses the two-qubit system to one of the sub-spaces
of the Bell basis.  For a detector with linear response this
corresponds to measurement of the total spin of the qubits. In the
other extreme of purely quadratic response the operator
$\sigma_y^1 \sigma_y^2 +\sigma_z^1\sigma_z^2$ is measured.
In both cases, collapse
naturally leads to spontaneous entanglement which can be identified by
measurement of the power spectrum and/or the average current of the
detector.
Asymmetry between the two qubits results in evolution between the
different measurement subspaces.
However, when the qubits are even weakly coupled
to the detector, a kind of quantum Zeno effect cancels the gradual evolution
and replaces it with rare, abrupt switching events.  We obtain the
asymptotic switching rates for these events and confirm them with
numerical simulations.  We show how such switching affects the
observable power spectrum on different time scales.
\end{abstract}
\pacs{73.23.-b; 03.65.Ta; 03.65.Xp; 03.67.Lx}

\maketitle

\section{Introduction}

The quantum Zeno effect is an intriguing prediction that arises in the
context of quantum measurement theory.\cite{BraginskyKhalili,Joos,Nakazato}
It states that a sequence of strong orthodox
measurements can ``freeze'' a system in its quantum state, so that in
the Zeno limit of very frequent measurements, the system is prevented from
decay and/or subsequent evolution.  Since its original
formulation,\cite{Khalfin,MisraSudarshan} there has been continued
theoretical development \cite{Peres,Joos1,QZE-development,another} as well as
experimental demonstration \cite{Itano,Fischer} in an ensemble of
quantum micro systems -- trapped ions (or atoms) probed via fast
(strong) quantum-optical measurements.

    With the recent experimental advances in the deliberate fabrication
of  two-level quantum systems, such as superconducting
qubits\cite{superqubits} or double quantum dots\cite{QDots} (DQD),
it has become possible to perform experiments on {\it individual}
quantum systems (see, e.g.,
Refs.\,\onlinecite{Kor-99-01},\onlinecite{Averin-SolSt}).
However, it can be very difficult to make repeated  
strong orthodox measurements on such systems.  Instead, a more practical
measurement scheme  utilizes a detector such as a quantum
point contact\cite{QDots} (QPC) or a single-electron
transistor\cite{SET} (SET) that is weakly coupled to the
qubit(s).
The measurement record in such
a situation is a fluctuating current $I(t)$ that accumulates a distinguishable
signal-to-noise ratio after some time.\cite{SET,Ilichev1}

    On one hand, the weak coupling between detector and qubit(s)
permits  the quantum system to remain relatively well isolated from
``outside'' classical noise. On the other hand, it means that instead
of simple abrupt collapse,\cite{vonNeumann} we have to deal with a theory
of continuous (weak) measurements of a single quantum system.
The development of such theories started long ago
\cite{Mensky,Gisin,Caves-86} and has attracted most attention in
quantum optics.\cite{Wiseman,Carmichael,Plenio}  Despite the similarity
of their underlying principles, the theories may differ significantly
in formalism and area of application; for solid-state qubits a ``Bayesian''
theory was developed relatively recently.\cite{Kor-99-01}  This formalism
is reviewed in Ref.\, \onlinecite{Kor-rev}, and its
equivalence to the quantum trajectory approach translated
\cite{Goan,Goan-2} from quantum optics is shown in
Ref.\, \onlinecite{Goan}.

    For single-qubit continuous measurement,\cite{Kor-osc,Makhlin} it was
shown that in a hypothetical situation of {\it very strong} continuous qubit-detector
coupling, it would be possible to approach the quantum Zeno limit.
For large (but finite) qubit-detector coupling
long periods of ``freezing'' of the single-qubit
state are interrupted by rare, abrupt jumps between qubit states.
Such jumps produce corresponding
jumps in the average detector current signal that
closely follows the qubit evolution.
This is in agreement with the classical character
of the strong coupling measurement regime.\cite{Kor-osc,ShepelyanskyDetect,RKMineq}

    In this paper we show that the Zeno-like regime can arise even in
a relatively practical {\it weak} continuous measurement context.  We
initially consider two identical qubits coupled symmetrically to a detector.
Measurement continuously collapses the two-qubit state into
{\it measurement subspaces}, i.e., stable ``points'' or
stable multi-dimensional subspaces of the two-qubit Hilbert space.
On a time scale
greater than the characteristic measurement time the collapse is like
a projective measurement in the Bell basis.
For a detector with linear or non-linear response\cite{RusKor-ent} this
corresponds to measurement of the total ``spin'' of the qubits
$(\overrightarrow{\sigma}^{(1)}+\overrightarrow{\sigma}^{(2)})^2$.
In the degenerate case of purely quadratic
detection\cite{prl-quadr} the
spin-1 subspace itself splits and effectively the operator
$\sigma_y^{(1)} \sigma_y^{(2)} +\sigma_z^{(1)} \sigma_z^{(2)}$ is
measured.  The Zeno-like regime manifests itself in a situation of
slightly asymmetric qubits, wherein the Hamiltonian attempts to evolve a
collapsed state out of its measurement subspace.  Our results show that
measurement stabilizes the state, stifling this evolution.  Instead of
gradual evolution, one finds rare quantum
switching events between different subspaces.

    Given the noisy experimental signal $I(t)$, such switching events
can be identified by consideration of the power spectrum and/or
average current.  Using the stochastic Bayesian approach we evaluate
the power spectrum in several important cases including the linear and
purely quadratic detection regimes.  We estimate the measurement time needed
to accumulate the spectrum and show that even for a non-ideal detector
it is possible to observe switching events if the lifetime of the
subspaces is large enough.  On a long time scale compared to the
subspace's lifetime, the switching events lead to an averaging of the
power spectrum (effective decreasing of the spectral peaks) and also
may produce a telegraph noise peak near zero frequency.  The resulting
long-time-average power spectrum is in accordance with both numerical
simulation of the spectrum through Bayesian equations\cite{RusKor-ent} and
with the ensemble averaged approach.\cite{prl-quadr,Mao}

   As expected in the quantum Zeno effect, the state ``survival
probability'' evaluated from quantum master equations
decays quadratically in time\cite{Khalfin,MisraSudarshan,Peres}
for small times and exponentially  for times\cite{Joos1} that are
sufficiently large compared to the measurement time.  We have obtained
analytically the corresponding ``lifetimes'' of the subspaces for
small deviations from the symmetric situation.  We have verified our
results by Monte Carlo simulations of the measurement process.

The presentation of our results is organized as follows.  In Sec. II,
we review the Hamiltonian and the Bayesian equations of motion for weak
continuous measurements of a two-qubit system.  Sec. III describes in
detail the measurement collapse scenarios that result from the
Bayesian equations.  In Sec. IV, the quantum Zeno effect is
demonstrated when the two qubits are asymmetrical.  Sec. V presents
conclusions.

\section{Model Hamiltonian and Bayesian equations of motion}

\subsection{Model Hamiltonian}

\begin{figure}
\vspace*{0.1cm}
\centering
\includegraphics[width=3.25in]{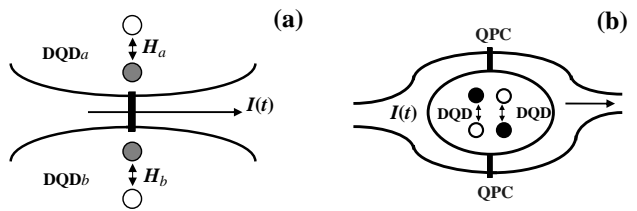}
\vspace{0.05cm}

\caption{Schematic of two qubits measured by a linear and non-linear detector.
(a): Realization based on double quantum dots measured by a QPC.
(b): Possible realization of
a purely quadratic detector based on double
quantum dots measured by two QPCs.}
\label{schematic}
\end{figure}

Fig. \ref{schematic}(a) shows possible realization of the setup that
we analyze.  Each qubit is made of a double quantum dot \cite{QDots}
(DQD), occupied by a single electron, while the detector is a quantum
point contact (QPC) located in between DQDs.  Another possible
realization (not shown) is based on single-Cooper-pair-box
(SCPB) qubits \cite{superqubits} measured by a
single-electron transistor (SET).

        In the Hamiltonian of the system,
\begin{equation}
  {\cal H} = {\cal H}_{QB} + {\cal H}_{DET} + {\cal H}_{INT} ,
\end{equation}
the first term describes two qubits alone with direct interaction
between them,
\begin{eqnarray}
{\cal H}_{QB}=
(\varepsilon _a/2)(a_\downarrow^\dagger a_\downarrow
                - a_\uparrow^\dagger a_\uparrow)
+H_a (a_\uparrow^\dagger a_\downarrow + a_\downarrow^\dagger a_\uparrow)
  &&\nonumber\\
+(\varepsilon _b/2)(b_\downarrow^\dagger b_\downarrow
             - b_\uparrow^\dagger b_\uparrow)
+H_b (b_\uparrow^\dagger b_\downarrow + b_\downarrow^\dagger b_\uparrow)
  &&\nonumber\\
+U(a_\uparrow^\dagger a_\uparrow-a_\downarrow^\dagger a_\downarrow)
(b_\uparrow^\dagger b_\uparrow-b_\downarrow^\dagger b_\downarrow) \qquad 
\label{H-2qb}
\end{eqnarray}
The amplitudes $H_a$ and $H_b$ describe tunneling within each qubit
while $\varepsilon_a$ and $\varepsilon_b$ are the energy biases of each qubit.
In the realization shown in Fig. \ref{schematic}(a),
$U$ gives the strength of the Coulomb interaction.  For other
realizations, the interaction may arise from a different physical
source.  In the absence of interaction ($U=0$), each qubit becomes an
independent two level system. The free Rabi
oscillation frequencies of qubits ``a'' and ``b'' are then given by
$\widetilde{\Omega}_a\equiv(4H_a^2+\varepsilon_a^2)^{1/2}/\hbar$ and
$\widetilde{\Omega}_b\equiv(4H_b^2+\varepsilon_b^2)^{1/2}/\hbar$
respectively.
In the following it will be convenient to define the Rabi frequencies at
zero energy bias $\Omega_a\equiv 2H_a/\hbar$ and
$\Omega_b\equiv 2H_b/\hbar$.

As an example of a detector we consider the case \cite{Gurvitz} of a
low transparency QPC, with Hamiltonian
\begin{equation}
{\cal H}_{DET}=\sum_l E_l c_l^\dagger c_l + \sum_r E_r c_r^\dagger c_r
+\sum_{l,r} (T c_r^\dagger c_l + \mbox{H.c.})
\label{detector}
\end{equation}
An operator of the form $c^\dagger_l$ ($c^\dagger_r$) creates an
electron in the left (right) lead of the detector.
The final term causes electrons to make transitions
between the leads.  We sum over all states $l$ in the left lead and
all states $r$ in the right lead.
The qubit--detector interaction term can be written as
\begin{eqnarray}
{\cal H}_{INT}=\sum_{l,r}  (a_\uparrow^\dagger a_\uparrow
         - a_\downarrow^\dagger a_\downarrow) (b_\uparrow^\dagger b_\uparrow
         - b_\downarrow^\dagger b_\downarrow)\,
         \chi\, c_r^\dagger c_l + \mbox{H.c.} &&
  \nonumber\\
+\sum_{l,r}  (a_\uparrow^\dagger a_\uparrow
         - a_\downarrow^\dagger a_\downarrow)\,
  \Delta T_a\, c_r^\dagger c_l + \mbox{H.c.} &&
  \nonumber\\
+\sum_{l,r}  (b_\uparrow^\dagger b_\uparrow
         - b_\downarrow^\dagger b_\downarrow)\,
  \Delta T_b\, c_r^\dagger c_l + \mbox{H.c.}.
\label{non-linear}
\end{eqnarray}
The amplitudes $T$ and $\Delta T_a$,
$\Delta T_b$, $\chi$
are assumed independent of the energy of
the tunneling electrons.
For simplicity, we will consider a situation when there is no
relative phase between these amplitudes.
Such a relative phase would lead to additional
decoherence.\cite{Kor-nonideal,Ruskov-osc,AverinSukhorukov}
From the form of (\ref{non-linear}), one sees that the detector's
measurement basis is $\mid \uparrow\rangle$, $\mid \downarrow \rangle$
for each qubit.  Eq. (\ref{non-linear}) is the most general
detector-system interaction in this basis in the case of two
qubits.\cite{prl-quadr}
Higher terms in Eq.\,(\ref{non-linear}) will appear for larger number of qubits.

Purely quadratic detectors ($\chi\neq 0$ and $\Delta T_{a,b}=0$)
may be of use to perform simple
quantum error correction protocols.\cite{AverinFazio} To realize
such a detector physically, one possibility is
to form two identical QPCs into two arms of an Aharonov-Bohm type
loop (no magnetic field is applied).  If two qubits are placed
exactly at the geometrically symmetric point of the loop as in
Fig. \ref{schematic}(b), we have a purely quadratic detector which
responds only on the relative state of the two qubits, i.e. it
does not distinguish between states
$\mid \uparrow_a \uparrow_b \rangle$ and $\mid \downarrow_a \downarrow_b \rangle$
and also does not distinguish between
$\mid \uparrow_a \downarrow_b \rangle$ and $\mid \downarrow_a \uparrow_b \rangle$.
Ref.\ \onlinecite{Mao} presented an explicit example of a non-linear
detector constructed to measure flux qubits
in the form of a superconducting loop interrupted by two Josephson junctions.

The four basis states of two qubits,
$|1\rangle \equiv \, \mid\uparrow_a \uparrow_b \rangle$,
$|2\rangle \equiv \, \mid\uparrow_a \downarrow_b \rangle$,
$|3\rangle \equiv \, \mid\downarrow_a \uparrow_b \rangle$,
$|4\rangle \equiv \, \mid\downarrow_a \downarrow_b \rangle$,
correspond to four values of the average current
through the detector:
\begin{eqnarray}
&&I_{1,2} = 2\pi (T\pm \chi +\Delta T_a \pm \Delta T_b)^2
\rho_l \rho_r e^2 V/\hbar \nonumber
\\
&&I_{3,4} = 2\pi (T\mp \chi -\Delta T_a \pm \Delta T_b)^2
\rho_l \rho_r e^2 V/\hbar  ,
\label{4currents}
\end{eqnarray}
where $V$ is the QPC voltage and $\rho_{l}$ and $\rho_{r}$ are the
densities of states in the left and right leads.

Note that if a very large voltage were applied to the QPC
for a short time, the current variances would become
much smaller than the corresponding  
current differences.
This is probably experimentally impractical, but in principle it
would constitute a strong orthodox measurement.

\subsection{Weak continuous measurements of qubits}

In the course of a measurement, our knowledge about a quantum system
is updated according to the result of the measurement.  In the case of
strong instantaneous measurements the update (``collapse'') is also
instantaneous.\cite{vonNeumann} In a weak continuous measurement, the
change of the system's quantum state occurs gradually.\cite{Kor-99-01}
The measured result is intrinsically noisy (it is a noisy
quasi-continuous current $I(t)$ or voltage $V(t)$ in the simplest
solid-state realizations), so that the information from the detector
is also acquired only gradually in time.

Quantitatively, such a random measurement process can be
described by Bayesian equations\cite{Kor-99-01}
for the system density matrix $\rho_{ij}(t)$
that take into account the actual measurement result $I(t)$.
For a system of qubits measured by a common
detector they can be written\cite{Kor-ent,Kor-nonideal}
as stochastic equations (in It\^{o} form\cite{Oksendal}):
\begin{eqnarray}
&& \dot{\rho}_{ij} =
[ I(t)-\sum_k \rho_{kk} I_k ]\,
\left(I_i+I_j-2\sum_k \rho_{kk}I_k \right)\, \frac{\rho_{ij}}{S_0}
        \nonumber \\
&&\qquad\ {}-\left[\frac{(I_i-I_j)^2}{4S_0} +\gamma_{ij}\right]\, \rho_{ij}
-\frac{\imat}{\hbar} [{\cal H}_{QB}, \rho]_{ij} ,
        \label{Bayes1}
\end{eqnarray}
where the summation is performed over all of the states of the system,
the $I_k$ are the corresponding current levels,
$S_0$ gives the background low-frequency noise in the detector, and $\gamma_{ij}$
is the decoherence rate.  For a 2-qubit system, there are $2^2 = 4$ states in the
sum over $k$ and
the $I_k$ are given by (\ref{4currents}).  Eq.\ (\ref{Bayes1}) assumes that
the detector is functioning in a weakly responding regime:
$|I_i - I_j| \ll I_i,I_j$; this justifies the use of a single low
frequency noise $S_0$ for all states $|k\rangle$ of the
system.

Eq.\ (\ref{Bayes1}) can be obtained by a general ``informational''
Bayesian approach \cite{Kor-99-01,Kor-ent,Kor-nonideal} (see also an analogous
derivation of the Bayesian equations for the measurement of a
nano-resonator\cite{Ruskov-osc}).
Since QPC is an ideal detector with
100\% quantum efficiency and no ``phase asymmetry'', the
evolution due to measurement basically reduces to the ``quantum
Bayes theorem.''\cite{GardinerZoller} It says
that at a sufficiently short time scale in which the evolution by $\cal{H}_{QB}$ can be
neglected:  1) evolution of the diagonal density matrix elements in the measurement
basis coincides with evolution of classical probabilities
described by usual Bayes formula and 2) the purity measure
$\rho_{ij}/\sqrt{\rho_{ii} \rho_{jj}}$ stays constant.
Such a derivation does not assume
linearity of the measurement device, so it is also relevant for the
general non-linear coupling of (\ref{non-linear}).
Essentially, Eq.\ (\ref{Bayes1}) describes the interplay between the
``intrinsic'' qubit evolution (Rabi oscillations) and the evolution
due to measurement. It implicitly assumes that the internal detector dynamics
of (\ref{detector}) and (\ref{non-linear}) is much faster than the qubit
dynamics due to $\cal{H}_{QB}$ and measurement.
This is seen explicitly in another
approach\cite{Kor-99-01,prl-quadr} that leads to the same Eq.\ (\ref{Bayes1}).
It uses generalized Bloch equations
that couple the qubit dynamics and the number $n$ of electrons that
have passed through the detector (for the one-qubit case see
Ref.\ \onlinecite{Gurvitz}).
One assumes that the typical detector decoherence time is much smaller
than the characteristic electron tunneling time, $\hbar/eV \ll e/I_0$
so that coherences between different passages of electrons
can be neglected.\cite{Kor-99-01}
In order for the current to be considered continuous, the characteristic
electron tunneling time must be much less than the time scale of the qubit
dynamics, $e/I_0 \ll 2\pi/\Omega$.  Then, in the weakly responding regime,
when individual tunneling events
produce only a small change of the qubit state it is convenient
to condition the evolution on the current $I(t)$.  These conditions are
sufficient for the stochastic process $\rho_{ij}(t)$ to be Markovian
and make it possible to obtain equation (\ref{Bayes1}).

To simulate individual realizations of the random measurement
process, the noisy detector current $I(t)$ can be expressed as
\begin{equation}
I(t) =\xi(t) + \sum_k \rho_{kk} I_k
\label{noisy-current} ,
\end{equation}
where $\xi (t)$ represents white noise
with spectral density $S_0$.
For a measurement with an ideal detector there are no individual qubit
decoherences ($\gamma_{ij}=0$).  As a result,
Eq.\,(\ref{Bayes1}) leads to purification of the qubit state in the
course of continuous measurement even though the qubit undergoes random time
evolution.\cite{Kor-99-01}

    Averaging over different realizations of the measurement process
$I(t)$ in Eq.\,(\ref{Bayes1}) means total ignorance of the particular measurement
result and is equivalent to ensemble averaging.
The ensemble average of the qubit state does exhibit decoherence
even for an ideal detector.
Averaging in Eq.\,(\ref{Bayes1}) is simply achieved by removing
the noise term.\cite{Oksendal}  The result is the standard
ensemble-averaged equations \cite{Leggett} with ensemble dephasing
rates $\Gamma_{ij} =(I_i-I_j)^2/4S_0 +\gamma_{ij}$
\begin{equation}
 \dot{\rho}_{ij} = -\Gamma_{ij} \, \rho_{ij} -\frac{\imat}{\hbar} [{\cal H}_{QB}, \rho]_{ij} .
\end{equation}
We see that even in the case $\gamma_{ij}=0$ the averaging has
produced a decoherence rate $(I_i-I_j)^2/4S_0$ which is
the quantum limited decoherence from an ``ensemble averaged'' point of
view.\cite{Kor-99-01}${}^{,}$\cite{Averin-SolSt,Pilgram,Clerk}

Individual qubit decoherence $\gamma_{ij} \ne 0$ will arise if the detector is not
ideal and may be understood as partial ignorance of the measurement result.\cite{Kor-99-01}
The contribution to $\gamma_{ij}$ is parameterized by the detector
ideality or efficiency
$\eta$ ($0\leq \eta \leq 1$)
\begin{equation}
\gamma_{ij} = (1/\eta -1)\, \frac{(I_i-I_j)^2}{4S_0}.
\label{eta}
\end{equation}
The partial ignorance may arise due to extra output noise\cite{Kor-nonideal}
or when the QPC detector has a finite temperature $T$.\cite{Kor-99-01}
For finite temperature, one takes into account the total current through the
QPC, $I(t)=I_{\rightarrow}(t) - I_{\leftarrow}(t)$, but ignores information contained in
the partial currents ``to the right'' and ``to the left'' in the detector.
Correspondingly the quantum ideality (efficiency) is reduced down to\cite{Kor-99-01,RusKor-sp}
$\eta=\tanh^2{(eV/2T)}$.\cite{ideality}
Notice that no mutual dephasing need be present between two states
$i,j$ if $I_i=I_j$. Indeed, in this case the detector cannot distinguish
between the two states and therefore there is no (relative) information to be received
and subsequently ignored. Then both $\gamma_{ij}$ and $\Gamma_{ij}$ vanish.
    Of course, additional qubit decoherence can arise
due to individual coupling of qubits to the environment.  The equation
for $\gamma_{ij}$ will then change as we will see in Sec. IV.

\section{Symmetric weak continuous measurements of two identical qubits}

In the quantum Zeno effect, measurement suppresses
Hamiltonian evolution,
repeatedly collapsing the system back to a given state.
To understand how this suppression happens, it is essential to develop
an understanding of the process of collapse.
In this section, we consider symmetric weak continuous measurement
of two identical qubits.
In this case the system collapses into {\it measurement subspaces} that can be
identified and characterized.
    We present a detailed analysis of the possible
collapse scenarios
for coupling to a linear, non-linear, or purely quadratic detection device.
This sets the stage for the appearance of quantum Zeno physics
in Sec. \ref{stabilization}.

\subsection{Preliminary view of dynamics}

To consider solutions of (\ref{Bayes1}), it is illuminating to
analyze first the dynamics of the Hamiltonian (\ref{H-2qb}) alone.
Introducing the Bell basis proves convenient:
\begin{eqnarray}
 |1\rangle^B \equiv (\mid \uparrow_a \downarrow_b \rangle
  -\mid \downarrow_a \uparrow_b \rangle )/\sqrt{2}, &&
\label{Bell1}\\
|2\rangle^B \equiv (\mid \uparrow_a \uparrow_b \rangle
   -\mid \downarrow_a \downarrow_b \rangle )/\sqrt{2}, &&
\label{Bell2}
\end{eqnarray}
\begin{eqnarray}
|3\rangle^B \equiv (\mid \uparrow_a \downarrow_b \rangle
   +\mid \downarrow_a \uparrow_b \rangle )/\sqrt{2}, &&
\label{Bell3}\\
|4\rangle^B \equiv (\mid \uparrow_a \uparrow_b \rangle
   +\mid \downarrow_a \downarrow_b \rangle )/\sqrt{2} \, .&&
\label{Bell4}
\end{eqnarray}
Using Eq.\ (\ref{H-2qb}) it is straightforward to
obtain the free evolution of the two qubits when
the qubits are unbiased, $\varepsilon_a=\varepsilon_b=0$, and
do not interact, $U=0$.  In this case,
the ``$+$" and ``$-$" subspaces of the Bell basis do not mix:
\begin{eqnarray}
&& |1\rangle^B \rightarrow \cos(\Delta\Omega t/2)\,|1\rangle^B -
i\sin(\Delta\Omega t/2)\,|2\rangle^B
\label{free-evol12}\\
&& |3\rangle^B \rightarrow \,\cos(\Omega t)\ |3\rangle^B -
i\sin(\Omega t)\ |4\rangle^B
\label{free-evol34} ,
\end{eqnarray}
where $\Delta\Omega\equiv\Omega_a-\Omega_b$ and $\Omega \equiv
(\Omega_a+\Omega_b)/2$.  This property persists
when the direct interaction is switched on
\begin{eqnarray}
&& |1\rangle^B \rightarrow
   [\cos(\Omega^{-} t) +
i\frac{U}{\hbar \Omega^{-}} \sin(\Omega^{-} t)]\,|1\rangle^B
\label{free-evolU12}\\
&&\qquad\qquad{} - i\frac{\Delta\Omega}{2\Omega^{-}}\sin(\Omega^{-}t)\,|2\rangle^B \\
&& |3\rangle^B \rightarrow
   [\cos(\Omega^{+} t) +
i\frac{U}{\hbar \Omega^{+}} \sin(\Omega^{+} t)]\,|3\rangle^B \nonumber\\
&&\qquad\qquad{} - i\frac{\Omega}{\Omega^{+}}\sin(\Omega^{+}t)\,|4\rangle^B \\
&& \Omega^{-} \equiv \sqrt{(\Delta\Omega/2)^2+(U/\hbar)^2}\\
&& \Omega^{+} \equiv \sqrt{\Omega^2+(U/\hbar)^2}
\label{free-evolU}.
\end{eqnarray}
Notice that $|1\rangle^B$ and $|2\rangle^B$ are eigenstates of
${\cal H}_{QB}$ if $\Delta \Omega = \Omega_a - \Omega_b =0$.  When the energy
biases $\varepsilon_a$, $\varepsilon_b$ are non-zero, there is no
simple solution to the free equations of motion.

\begin{figure}
\vspace*{0.2cm}
\centering
\includegraphics[width=3.35in]{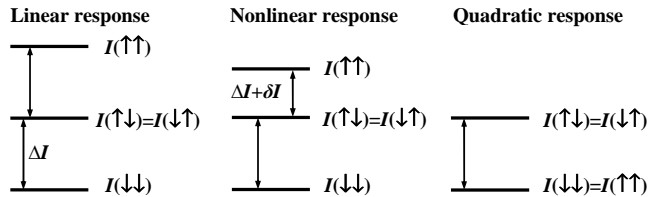}
\vspace{0.2cm}

\caption{Current levels for symmetrically coupled qubits,
$\Delta I_a = \Delta I_b = \Delta I$.
Linear, non-linear, and purely quadratic response.}
\label{levels}
\end{figure}

Turning to the coupling Hamiltonian (\ref{non-linear}), we note that
our assumption of symmetric qubit couplings
implies $\Delta T_a =\Delta T_b$.  In this case, two currents in (\ref{4currents})
coincide, $I_2=I_3\equiv I_{23}$, as seen in Fig.\,\ref{levels}, so the measurement cannot
distinguish between the states $|2\rangle\!\!\!\!=\mid\uparrow_a \downarrow_b \rangle$ and
$|3\rangle\!\!\!=\mid\downarrow_a \uparrow_b \rangle$.
Qualitatively, the measurement alone tends to continuously collapse the system
towards one of the subspaces with definite current.
For a linear detector ($\chi = 0$ in Eq.\,(\ref{non-linear})), there are three
possible values of current $I_1$, $I_4$, $I_{23}$.  Therefore, measurement collapses
the system toward $|1\rangle$ or $|4\rangle$ or the subspace \{$|2\rangle$, $|3\rangle$\}.
The current values are evenly spaced $I_1 - I_{23} = I_{23} - I_4$.
For a non-linear detector ($\chi\neq 0$ in (\ref{non-linear})), there are again
three possible currents, so that measurement again collapses the system toward the same
three subspaces. However, the currents are not evenly spaced.
For a purely quadratic detector, ($\Delta T_{a,b} = 0$ in (\ref{non-linear})
while $\chi \ne 0$), $I_1=I_4$ and $I_2 = I_3$, so the measurement itself collapses the system
toward the subspace  \{$|1\rangle$, $|4\rangle$\} or toward the subspace \{$|2\rangle$, $|3\rangle$\}.

To describe the measurement process in more detail, it is convenient to
introduce dimensionless qubit-detector couplings
${\cal C}_a  \equiv  \hbar (\Delta I_a)^2/S_0H_a$ and  ${\cal C}_b\equiv \hbar(\Delta I_b)^2/S_0H_b$
where the current differences are defined as
$\Delta I_a\equiv I_2-I_4$, $\Delta I_b \equiv I_3-I_4$, and
$\Delta I\equiv(\Delta I_a+\Delta I_b)/2$;
weak coupling corresponds to ${\cal C}_a,{\cal C}_b \lesssim 1$.
To characterize the non-linearity of the detector, it is useful to define
\begin{equation}
\delta \equiv \frac{\delta I}{\Delta I} =
 \frac{[(I_1 - I_3) - (I_2 - I_4)]}{[(I_2+I_3)/2 - I_4]}
\label{non-linearity},
\end{equation}
where $\delta I \equiv (I_1 - I_3) - (I_2 - I_4)$.
These definitions apply in general to the non-symmetric coupling case;
for identical symmetrically coupled qubits,    
$\Delta I_a = \Delta I_b = \Delta I$ and ${\cal C}_a = {\cal C}_b \equiv {\cal C}$,
and $\delta I$ is indicated on Fig.\ref{levels}.
For linear coupling $(\chi = 0)$ one has $\delta  = 0$ while
for the purely quadratic case $\delta = -2$.

\subsection{Bayesian equations in the Bell basis}

It is instructive to rewrite the Bayesian equations  (\ref{Bayes1}) in the Bell basis
(\ref{Bell1}) - (\ref{Bell4})
\begin{equation}
\frac{d \rho_{ij}^B}{dt} = \left.\frac{d \rho_{ij}^B}{dt}\right|_{meas}  +
\left.\frac{d \rho_{ij}^B}{dt}\right|_{H-evol}
\label{BayesBell} ,
\end{equation}
where we distinguish a non-unitary part due to measurement from a unitary part
due to free Hamiltonian evolution.
For qubits with identical couplings,
these contributions  (in It\^{o} form) are:
\begin{eqnarray}
\lefteqn{ \left.\frac{d \rho_{ij}^B}{dt}\right|_{meas}=
\xi_{ij} \mbox{-term} - \frac{(\Delta I)^2}{\eta\, 4S_0}\times \qquad\qquad\qquad  }
&&   \label{Bayesevol}  \\
&&  \left[ \left(
\begin{array}{cccc}
  0    & \rho_{12}^B                  &   0         & \rho_{14}^B \\
\cdot  & \frac{\rho_{22}^B-\rho_{44}^B}{2}\,(2+\delta)^2& \rho_{23}^B & \imat\, \mbox{Im} \rho_{24}^B\,(2+\delta)^2 \\
 \cdot & \cdot                        &   0         & \rho_{34}^B \\
\cdot  & \cdot                        & \cdot       &\frac{\rho_{44}^B-\rho_{22}^B}{2}\,(2+\delta)^2
\end{array}
\right)  \right. \nonumber \\
&& \left. + (\delta + \frac{1}{2}\delta^2) \,\left(
\begin{array}{cccc}
  0    & \rho_{12}^B + \rho_{14}^B &   0                     & \rho_{12}^B + \rho_{14}^B \\
 \cdot & 0                         & \rho_{23}^B+\rho_{43}^B & 0 \\
 \cdot & \cdot                     &   0                     & \rho_{32}^B + \rho_{34}^B \\
 \cdot & \cdot                     & \cdot         & 0
\end{array}
\right) \right].  \nonumber
\end{eqnarray}
The entries indicated with a dot are determined by the hermiticity of the matrices.

The noise term itself reads
\begin{eqnarray}
\lefteqn{\xi_{ij} \mbox{-term}=
 - \frac{2\,\Delta I}{S_0}\, \xi(t)\,\times
\qquad\qquad\qquad } &&  \label{Bayesevolxi} \\
&& \left[ \left(
\begin{array}{cccc}
z \rho_{11}^B & z \rho_{12}^B +\frac{\rho_{14}^B}{2} & z \rho_{13}^B                       &  z \rho_{14}^B -\frac{\rho_{12}^B}{2}\\
\cdot         & z (\rho_{22}^B -\frac{1}{2})         & z \rho_{23}^B -\frac{\rho_{43}^B}{2} & z \rho_{24}^B -\frac{\rho_{22}^B+\rho_{44}^B}{2} \\
\cdot         &\cdot                                 & z \rho_{33}^B            & \ \, z \rho_{34}^B -\frac{\rho_{23}^B}{2}  \\
\cdot         &\cdot                                 & \cdot        & z (\rho_{44}^B -\frac{1}{2})
\end{array}
\right) \right.  \nonumber \\
&& \left. - \frac{\delta}{4}\,\left(
\begin{array}{cccc}
  0     & \rho_{12}^B + \rho_{14}^B  &   0                     & \rho_{12}^B + \rho_{14}^B \\
 \cdot  & \rho_{22}^B-\rho_{44}^B    & \rho_{23}^B+\rho_{43}^B &2 \rho_{24}^B +(\rho_{22}^B+\rho_{44}^B) \\
 \cdot  & \cdot                      &   0                     & \rho_{32}^B + \rho_{34}^B \\
 \cdot  & \cdot                      & \cdot         &\!\!\!\!\!\!-\,(\rho_{22}^B-\rho_{44}^B)
\end{array}
\right) \right]  \nonumber ,
\end{eqnarray}
where it proves convenient to introduce the quantity
\begin{equation}
z \equiv \rho_{11}(1+\delta) - \rho_{44} = 2\mbox{Re}\rho_{24}^B \left(1+\frac{\delta}{2}\right)
 + \frac{\delta}{2}\,\left(\rho_{22}^B+\rho_{44}^B\right)
\label{zee}
\end{equation}
given in the measurement basis and in the Bell basis.
Since the detector signal (\ref{noisy-current}) can be
expressed, using  (\ref{non-linearity}) and (\ref{zee}) as,
\begin{equation}
I(t) = I_{23}+ \Delta I\,z(t) + \xi(t)
\label{noisy-curr-z} ,
\end{equation}
we see that $z(t)$ gives the two-qubit signal.
In addition, for linear coupling $z(t)$  
is equal to the average $z$-component of the total spin.

The unitary part has one contribution proportional to the
Rabi frequency $\Omega$, one proportional to the energy bias $\varepsilon$,
and one proportional to the interqubit coupling $U/\hbar$:
\begin{equation}
\left. \frac{d \rho_{ij}^B}{dt}\right|_{H-evol} =
\left.\frac{d \rho_{ij}^B}{dt}\right|_{\Omega} + \left.\frac{d \rho_{ij}^B}{dt}\right|_{\varepsilon}
+ \left.\frac{d \rho_{ij}^B}{dt}\right|_{U}
\end{equation}
where
\begin{equation}
\left.\frac{d \rho_{ij}^B}{dt}\right|_{\Omega}=\Omega
\left(
\begin{array}{c@{\:}c@{\:}c@{\:}c}
0    \ \     & 0      \     & \imat \rho_{14}^B  \         & \imat \rho_{13}^B \\
\cdot\ \     & 0      \     & \imat \rho_{24}^B  \         & \imat \rho_{23}^B \\
\cdot\ \     & \cdot  \     &-2\,\mbox{Im}\rho_{34}^B\ & \imat (\rho_{33}^B-\rho_{44}^B)\\
\cdot\ \     & \cdot  \     & \cdot        \           &2\,\mbox{Im}\rho_{34}^B
\end{array}
\right),
\label{H-evol-om}
\end{equation}
\begin{equation}
\left.\frac{d \rho_{ij}^B}{dt}\right|_{\varepsilon}=\frac{\varepsilon}{\hbar}
\left(
\begin{array}{c@{\:}c@{\:}c@{\:}c}
0    \   & \imat \rho_{14}^B  \     & 0  \            & i \rho_{12}^B \\
\cdot\   &-2\,\mbox{Im}\rho_{24}^B\ &-i \rho_{43}^B\  & i (\rho_{22}^B-\rho_{44}^B) \\
\cdot\   & \cdot\                   & 0  \            & i\rho_{32}^B \\
\cdot\   & \cdot\                   & \cdot\          & 2\,\mbox{Im}\rho_{24}^B
\end{array}
\right),
\label{H-evol-eps}
\end{equation}
and
\begin{equation}
\left.\frac{d \rho_{ij}^B}{dt}\right|_{U}=2 \frac{U}{\hbar} \
\left(
\begin{array}{c@{\:}c@{\:}c@{\:}c}
0    \ \    & i \rho_{12}^B & 0    \         & i \rho_{14}^B \\
\cdot\ \    & 0             &-i \rho_{23}^B\ & 0             \\
\cdot\ \    & \cdot\        & 0    \         & i \rho_{34}^B \\
\cdot\ \    & \cdot\        & \cdot\         & 0
\end{array}
\right).
\label{H-evol-U}
\end{equation}

The state (\ref{Bell1}) which we have denoted $|1\rangle^B$ has spin-0,
while the other three Bell states (\ref{Bell2}) - (\ref{Bell4})
comprise the spin-1 subspace.
By inspecting Eqs. (\ref{Bayesevol})
and (\ref{Bayesevolxi}), one sees that measurement does not mix
the spin-0 and spin-1 subspaces.
Indeed, when the system is in the spin-0 subspace, $\rho_{11}^B=1$,
then $d\rho_{11}^B/dt$ vanishes (since $z=0$). It also vanishes when the system
is in the spin-1 subspace ($\rho_{11}^B=0$).
Since $\mbox{Tr}\rho^B=1$, it follows that if the system is
initially in the spin-1 subspace it will remain there.
In the purely quadratic measurement case ($\delta=-2$) we have from (\ref{zee})
$z=-(\rho_{22}^B+\rho_{44}^B)$, and calculation shows that the spin-0 state
$|1 \rangle^B$, the state $|2\rangle^B$, and the ``+'' subspace,
\{$|3\rangle^B,|4\rangle^B$\}, similarly do not mix.  For unbiased qubits,
$\varepsilon =0$, Hamiltonian evolution via Eqs. (\ref{H-evol-om})
and (\ref{H-evol-U}) also
neither mixes the spin-0 and spin-1 subspaces nor the three subspaces of the
purely quadratic regime.
Gradual collapse to one of these subspaces occurs during the qubit evolution
as numerical simulation shows.
In the linear and non-linear cases there are two
collapse scenarios: collapse to the spin-0 or to the spin-1 subspace, so that
the operator $(\overrightarrow{\sigma}^{(1)} + \overrightarrow{\sigma}^{(2)})^2$
is being effectively measured.  In the purely quadratic case, there are three collapse
scenarios; we can say that we are effectively measuring  the operator
$\sigma_y^{(1)} \sigma_y^{(2)} +\sigma_z^{(1)} \sigma_z^{(2)}$,
whose eigenspaces are $|1\rangle^B$, $|2\rangle^B$,
and the ``+'' subspace \{$|3\rangle^B,|4\rangle^B$\} with eigenvalues
$-1$, $1$ and $0$ respectively.
One can confirm using (\ref{BayesBell}) that the measurement subspaces
associated with these collapse scenarios are attractive.
In other words, small deviations from a collapsed state tend toward
zero under the evolution (\ref{BayesBell}).
Similar attraction was obtained previously in an analysis of a sequence of
strong measurements.\cite{RusKor-ent}

We characterize below the collapse scenarios
by calculating the probability of collapse, typical collapse time,
and power spectrum of the detector current.

\subsection{Collapse scenarios when $\epsilon = 0$ and $U=0$}

\subsubsection{Linear and non-linear detection}

Examination of ensemble averaged equations
(master equations) provides insight into the collapse scenarios.
These equations can be obtained from (\ref{BayesBell}) by simply
eliminating the noise term (\ref{Bayesevolxi}).  For
both linear and non-linear detection, the master equations have the
following stationary solution
\begin{eqnarray}
&&\rho_{ij,st}^B=0, \qquad\quad \mbox{for}\ i\neq j
 \nonumber\\
&&\rho_{11,st}^B=\rho_{11}^B(0),
 \nonumber\\
&&\rho_{22,st}^B=\rho_{33,st}^B=\rho_{44,st}^B=\frac{1}{3}[1-\rho_{11}^B(0)]
\label{stationary-lin}.
\end{eqnarray}
The solution $\langle \rho_{11}^B (t) \rangle$ of the ensemble averaged
equations is a constant in time, $\langle \rho_{11}^B (t) \rangle= \rho_{11}^B (0)$.
Taking into account     
that after some measurement time\cite{objectivity} $t\gg\tau_{meas}$
all members of the ensemble eventually collapse to
$\rho_{11}^B(t)=1$ or $\rho_{11}^B(t)=0$
it must be that the fraction of members that
collapse to $\rho_{11}^B(t) = 1$
is equal to $\langle \rho_{11}^B (t) \rangle$. Thus,
one expects that the probability of collapse
to the spin-0 subspace is $\rho_{11}^B(0)$, while the probability
of collapse to the spin-1 subspace is $1-\rho_{11}^B(0)$.

\begin{figure}
\centering
\includegraphics[width=3.1in]{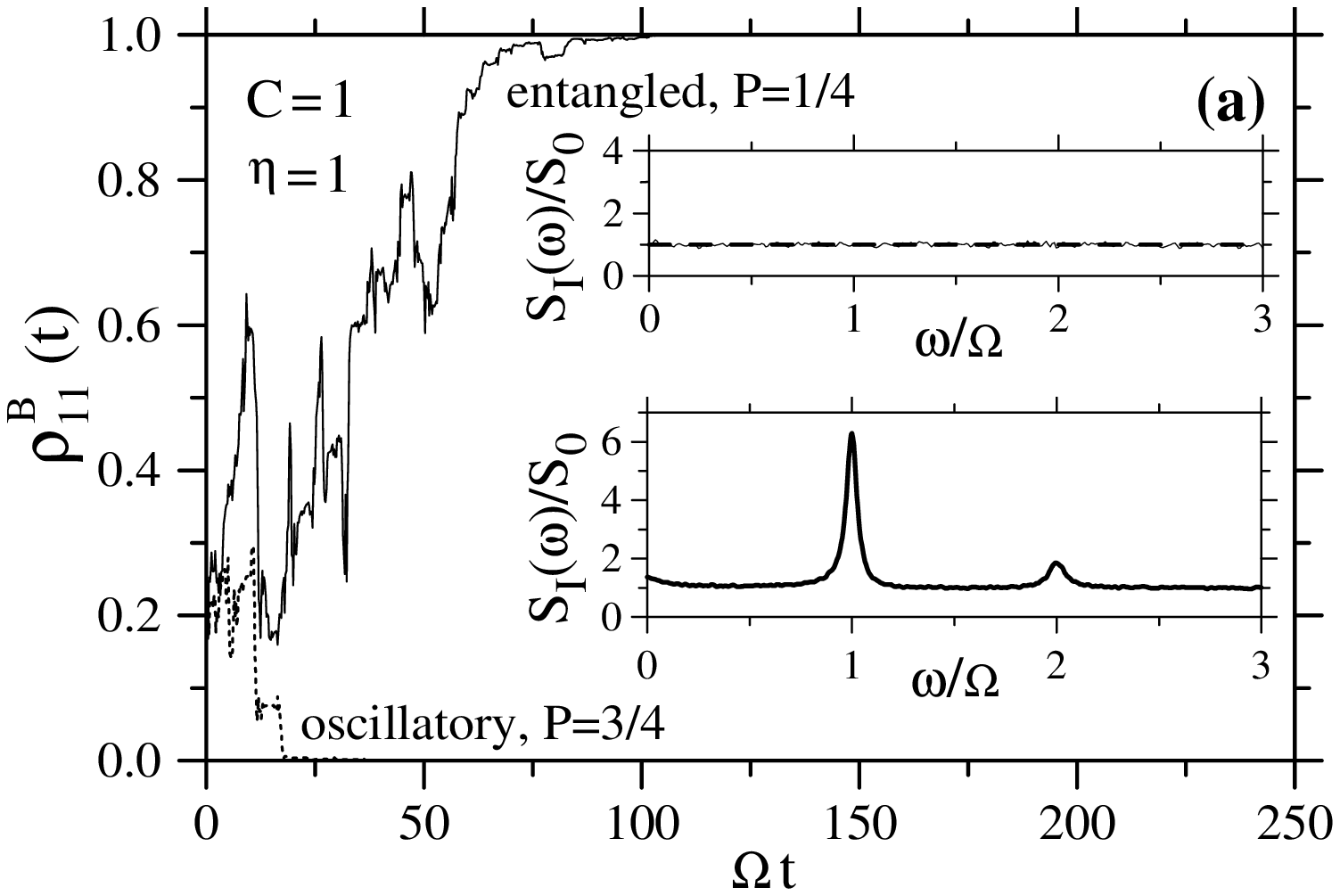}
\centering
\includegraphics[width=3.1in]{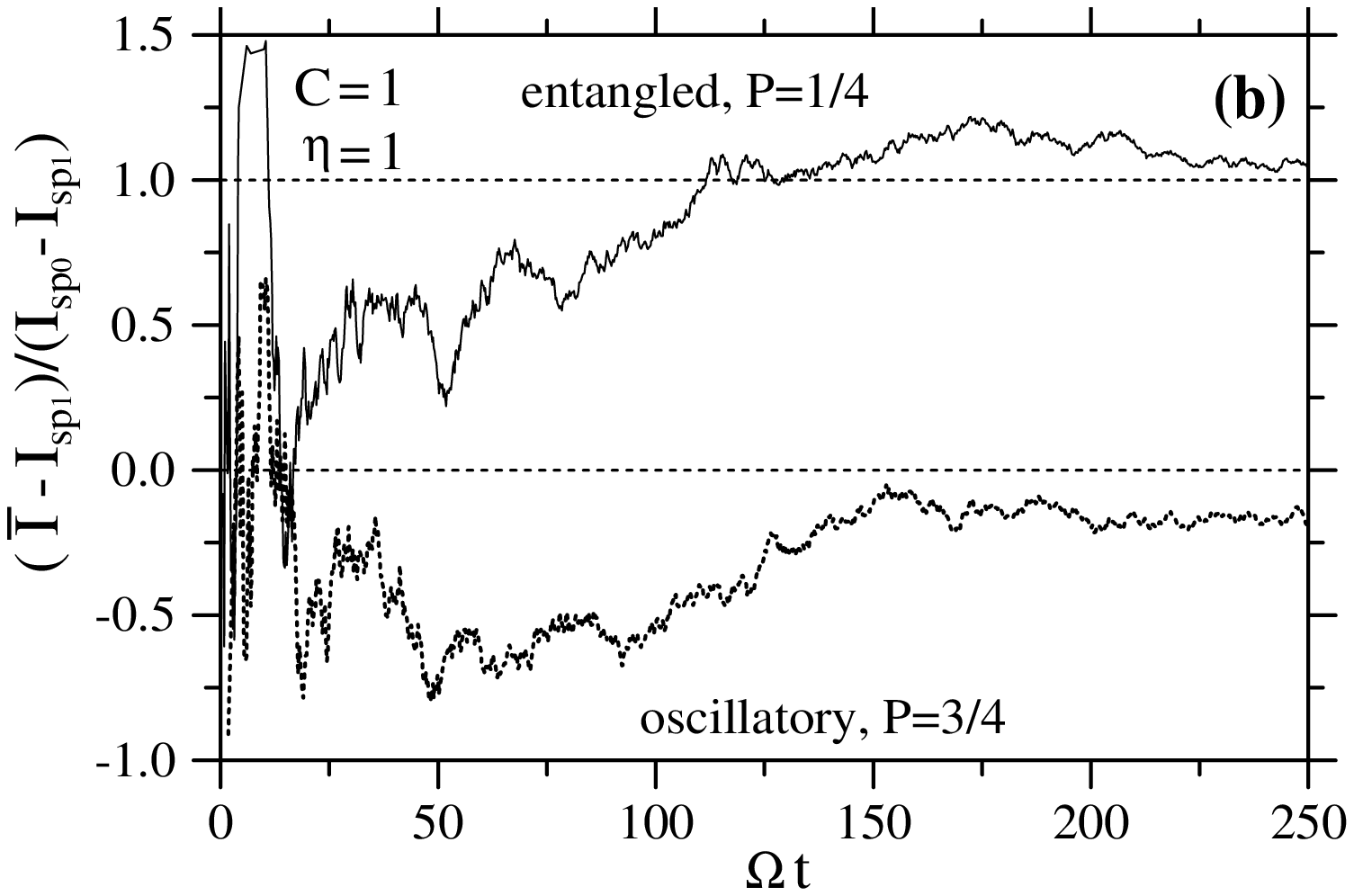}
\vspace{0.2cm}
\caption{ Two numerical realizations of  two-qubit evolution starting
from the fully mixed state for a {\it non-linear measurement}, $\delta=-1$,
by an ideal ($\eta = 1$) detector.
(a) Evolution of $\rho_{11}^B$. The upper (solid) line illustrates the scenario
of collapse into the entangled Bell state $|1\rangle^B$, while the lower
(dotted) line
shows a collapse into the orthogonal subspace. The two insets show the
corresponding spectral densities $S_I(\omega )$ of the detector current.
(b) The detector current $\overline{I}$ averaged over the whole time
starting from $t=0$ is shown for the same collapse scenarios as in (a).
Dashed horizontal lines correspond to the normalized averaged currents
$I_{sp_0}$, $I_{sp_1}$.
 }
\label{fig-collapse}
\end{figure}

Extensive Monte Carlo
simulations of (\ref{BayesBell}) were performed for weak coupling (${\cal C}$ between
$1/4$ and $1$).  As illustrated in Fig.\ \ref{fig-collapse}, they show that any initial
state either collapses eventually into the spin-0 state or collapses
to the orthogonal, spin-1 subspace within which it performs oscillations.
Within a level of accuracy consistent with the number of trials, the
probability of evolving into the Bell state $|1\rangle^B$ coincides
with the initial value $\rho_{11}^B(0)$. In particular, starting from
a completely mixed state the probability is $\rho_{11}^B(0)=1/4$.

\begin{figure}
\centering
\includegraphics[width=2.85in]{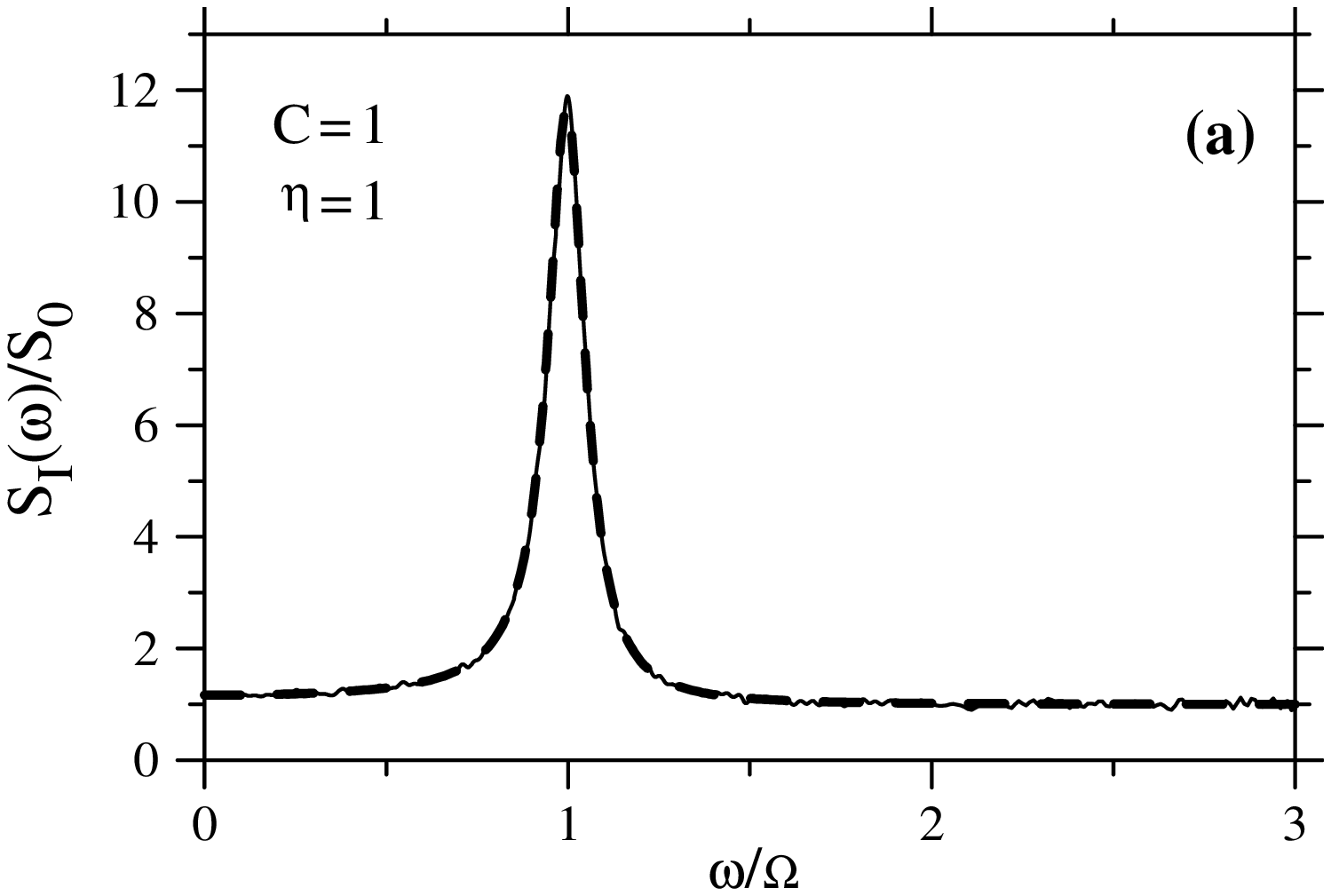}
\vspace{0.1cm}

\centering
\includegraphics[width=2.85in]{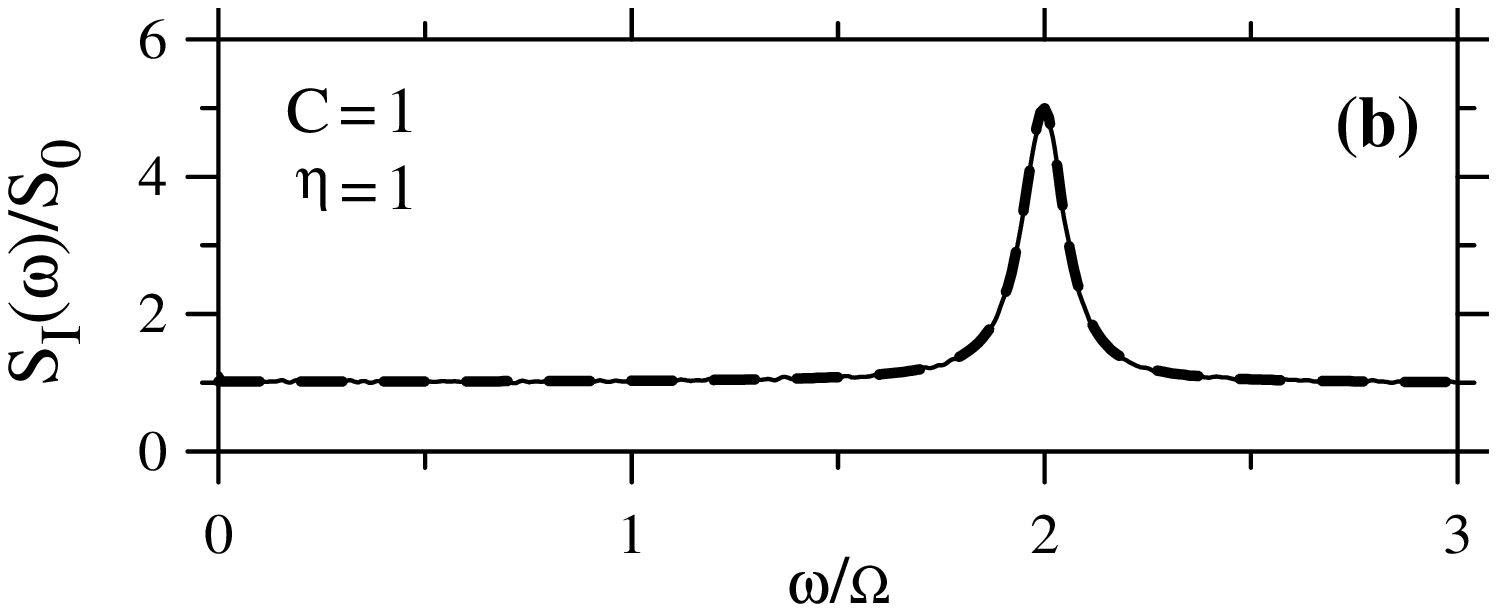}
\vspace{0.1cm}
\caption{ The two-qubit power spectrum in the oscillating scenario
in the linear (a) and purely quadratic (b) detection regimes.
Analytical calculation (dashed line), given by
Eqs. (\ref{spectrum-lin}),(\ref{spectrum-quadr})
practically coincides with numerics.
 }
\label{fig-spectr}
\end{figure}

     We can monitor the process of
collapse of $\rho_{ij}(t)$  through the
Bayesian equation (\ref{Bayes1}) for a given  measurement
result $I(t)$.
Collapse is also revealed  in the appearance of different
power spectra for $I(t)$
depending upon the final state (insets in Fig.\ \ref{fig-collapse}(a)).
In the case of collapse to the 
entangled state $|1\rangle^B$,
the spectrum is flat and coincides with the shot noise spectral density
$S_0$.  In the case of collapse to the spin-1 subspace, the qubit
performs oscillations within the subspace.  In the linear case the
oscillations lead to a single peak in the power spectrum
(Fig.\,\ref{fig-spectr}(a)) at the Rabi frequency $\Omega$ with peak
height $(32/3)S_0$ and width $\Gamma=(\Delta I)^2/4S_0$
that is confirmed analytically below.
In the general non-linear case, one sees peaks in the power spectrum at the
Rabi frequency $\Omega$ and also at zero and double the Rabi
frequency (Fig.\, \ref{fig-collapse}(a), lower inset).

The average detector current using (\ref{noisy-current}) is
$\langle I(t) \rangle = \sum_k I_k \langle \rho_{kk}(t) \rangle_t$.
After collapse, the time average $\langle \rho_{kk}(t) \rangle_t$
can be calculated using the stationary ensemble averaged solution
(\ref{stationary-lin})
-- the initial transient period of the order of the measurement time can be neglected.
In the spin-0 scenario one finds $\langle \rho_{22}(t) \rangle_t=\langle \rho_{33}(t) \rangle_t=1/2$,
implying $\langle I(t)\rangle =I_{sp_0} = (I_2 + I_3)/2 = I_{23}$.
In the spin-1 scenario
$\langle \rho_{11}(t) \rangle_t=\langle \rho_{44}(t) \rangle_t=1/3$ and
$\langle \rho_{22}(t) \rangle_t=\langle \rho_{33}(t) \rangle_t=1/6$, so that
$\langle I(t) \rangle = I_{sp_1} =(I_1+I_{23}+I_4)/3$.
In the linear case (Fig.\,\ref{levels}),
the average current in the detector is the same in both scenarios
($\langle I(t) \rangle=I_{sp_0}=I_{sp_1}=I_{23}$)
since $I_{23} = (I_1 + I_4)/2$. Therefore the spin-0 and spin-1
subspaces are indistinguishable by average currents and we
have to measure the power spectrum  in order to distinguish them.
In Appendix B we have evaluated the typical time needed to accumulate
the spectral peak; it is on the order of $1/\Gamma \eta^2$.

In the non-linear case, the final
state can be identified by the average current in the detector:
$I_{sp_0}=I_{23}$ in the spin-0 scenario and $I_{sp_1}=(I_1+I_{23}+I_4)/3$
in the oscillating spin-1 case (Fig.\ \ref{fig-collapse}(b)).
Thus one expects  the current signal averaged over
sufficiently long times $\Delta t \gtrsim 1/\Omega$,
$\overline{I}=(1/\Delta t) \int_0^{\Delta t} I(t') dt'$,  to be Gaussian distributed
with average
either $I_{23}$ or $(I_1+I_{23}+I_4)/3$ and with variance
$S_0/2\Delta t$.
We have confirmed this conjecture by numerical simulations of the detector current.
As a result, the typical time necessary to distinguish between the two scenarios with
a signal-to-noise ratio of 1 is on the order \cite{Kor-99-01} of
$\tau_{meas} = 2 S_0/(I_{sp_0}-I_{sp_1})^2=18 S_0/(2I_{23}-I_1-I_4)^2$.
This estimate is self-consistent since we are working in the weak coupling regime
($\Gamma \ll \Omega$), so that the measurement time $\tau_{meas} \gg 1/\Omega$.
Hence, for $\Delta t \sim \tau_{meas}$, the current $\overline{I}$ in the oscillating
scenario is effectively averaged to the value $I_{sp_1}$ over many periods of oscillation.

\begin{figure}
\centering
\includegraphics[width=2.9in]{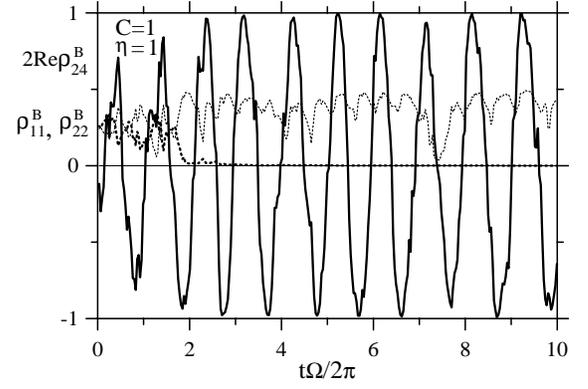}
%
\caption{ Linear weak ideal measurement.
State purification in case of oscillating
spin-1 collapse scenario ($\rho_{11}^B \rightarrow 0$,
thick dashed line) starting from a fully mixed state.
The qubits signal $z=2\mbox{Re}\rho_{24}^B$, Eq.\,(\ref{zee})
establishes quantum oscillations at Rabi frequency
(thick solid line) with slowly fluctuating amplitude and phase.
Also shown are  fluctuations of $\rho_{22}^B$ (thin dotted line).
 }
\label{fig-pur-osc}
\end{figure}

Both in linear and non-linear cases when collapse to $|1\rangle^B$ occurs,
purification of the state is
clearly inevitable even for a non-ideal detector with $\eta < 1$.
Simulation (see Fig.\, \ref{fig-pur-osc}) shows that for an ideal detector,
the qubit state purifies also in the case of collapse to the spin-1
subspace. (Such purification during oscillation in the weak
measurement regime is similar to the purification in the one-qubit
case;\cite{Kor-99-01} for analogous purification in the case of
measurement of an oscillator, see Ref.\,\onlinecite{Ruskov-osc}).
After initial relaxation, the surviving nondiagonal matrix elements in
the Bell basis are found numerically to satisfy
\begin{eqnarray}
(\mbox{Re}\rho_{24}^B)^2=\rho_{22}^B\rho_{44}^B,\quad
(\mbox{Im}\rho_{23}^B)^2=\rho_{22}^B\rho_{33}^B,
&&\nonumber\\
(\mbox{Im}\rho_{34}^B)^2=\rho_{33}^B\rho_{44}^B  \qquad &&
\label{nonzeros} \\
\mbox{Im}\rho_{24}^B=\mbox{Re}\rho_{23}^B=\mbox{Re}\rho_{34}^B=0 &&
\label{zeros}
\end{eqnarray}
which imply that the pure-state condition $\rho^2 = \rho$ is satisfied
by the matrix $\rho$.

In the weak coupling case ${\cal C}\lesssim 1$ for a linear ideal detector
($\eta=1$) it is possible to characterize further the evolution of the
density matrix.  Since purification of the state leads to Eqs. (\ref{nonzeros}) and (\ref{zeros}),
of the $8$ real degrees of freedom in
the density matrix of the collapsed system, only $2$ remain
independent.  The system oscillates with slowly fluctuating amplitude
and phase (see Fig. \ref{fig-pur-osc}).  Taking into account the free
Hamiltonian evolution, Eq.\,(\ref{H-evol-om}), we can write
\begin{eqnarray}
z = 2\mbox{Re}\rho_{24}^B=A(t)\cos (\Omega t +\varphi (t)), \qquad\qquad\quad\, &&
\label{z-parameter}\\
y \equiv 2\mbox{Im}\rho_{23}^B=A(t)\sin (\Omega t+\varphi (t)), \qquad\qquad\quad\, &&
\label{y-parameter}\\
z_1\equiv\rho_{33}^B-\rho_{44}^B=-B(t)\cos (2\Omega t +2\varphi (t)), \qquad\ &&
\label{z1-parameter}\\
y_1\equiv 2\mbox{Im}\rho_{34}^B=-B(t)\sin (2\Omega t +2\varphi (t)). \qquad\quad\ &&
\label{y1-parameter}
\end{eqnarray}
Here, $A(t)=2\,\sqrt{\rho_{22}^B(1-\rho_{22}^B)}$ and
$B(t)=1-\rho_{22}^B$, so it is natural to regard $\rho_{22}^B(t)$ and $\varphi(t)$
as the two independent variables.  The quantities $A(t)$, $B(t)$, and $\varphi(t)$ are
constants in the absence of measurement while in the presence of measurement they
slowly fluctuate in time.
(The stationary distribution of $\rho_{22}^B$ is calculated in Appendix A.)
Eqs. (\ref{z-parameter})-(\ref{y1-parameter})
complete the description
of the collapse subspaces in the case of weak linear ideal measurements.

\subsubsection{Purely quadratic detection}
\label{pqd}

In the case of purely quadratic measurement ($I_1=I_4\equiv I_{14}$, $\delta =-2$),
the stationary solution of the master equation is
\begin{eqnarray}
&&\rho_{ij,st}^B=0, \qquad\quad \mbox{for}\ i\neq j
 \nonumber\\
&&\rho_{11,st}^B=\rho_{11}^B(0),\quad \rho_{22,st}^B=\rho_{22}^B(0)
 \nonumber\\
&&\rho_{33,st}^B=\rho_{44,st}^B=\frac{1}{2}
[1-\rho_{11}^B(0)-\rho_{22}^B(0)]
\label{stationary-quadr} .
\end{eqnarray}
This suggests that with probability $\rho_{11}^B(0)$ the system collapses under
measurement to $|1\rangle^B$, with probability $\rho_{22}^B(0)$
the system collapses to $|2\rangle^B$, and with probability
$1-\rho_{11}^B(0)-\rho_{22}^B(0)$ the system collapses to the remaining
subspace.

Numerical simulations verify these three collapse scenarios.  The
average currents are $I_{23}$, $I_{14}$, and $(I_{23}+I_{14})/2$ for
the three scenarios so that the current spacing is $\Delta I/2$.
In the case of collapse to the two-dimensional
\{$|3\rangle^B,|4\rangle^B$\} subspace, the two-qubit system
performs quantum oscillations as seen in Eq.\,(\ref{free-evol34}).
At times larger than the oscillation period
(that are relevant in the weak measurement regime)
we can estimate a typical time necessary to distinguish
between the three scenarios that is on the order of
$\tau_{meas} \approx 8S_0/(\Delta I)^2$.

We checked numerically that the probabilities of the three collapse
scenarios are given by the initial values $\rho_{11}^B(0)$,
$\rho_{22}^B(0)$, and $1-\rho_{11}^B(0)-\rho_{22}^B(0)$ respectively.
In the case of collapse to the two-dimensional subspace, the qubits
oscillate, producing one peak in the power spectrum at double the Rabi
frequency (Fig.\,\ref{fig-spectr}(b)), with peak height $4 S_0$ and
width $\Gamma=(\Delta I)^2/4S_0$.

It is interesting to note the difference between the weak measurement
results presented here and the hypothetical (experimentally difficult)
strong measurement case. A strong, purely quadratic measurement was suggested
in Ref. \onlinecite{AverinFazio} as a means of measuring the relative
state of two qubits for use in error correction procedures.  In the strong
measurement case the qubit Hamiltonian dynamics is much slower than
the measurement dynamics -- the measurement occurs within a time much
shorter than the qubit oscillation period
$\tau_{meas}\sim S_0/(\Delta I)^2\ll 2\pi/\Omega$.
As a result, the two-qubit system collapses to subspaces with
average current $I_{14}$ or $I_{23}$, corresponding to ``parallel''
or ``anti-parallel'' qubits (Fig.\,\ref{levels}), given in the Bell basis
by \{$|2\rangle^B,|4\rangle^B$\} and \{$|1\rangle^B,|3\rangle^B$\}.
Therefore, effectively, the operator $\sigma_z^{(1)} \sigma_z^{(2)}$ is measured.
In the weak measurement situation considered in this paper,
the measurement time is much longer than the qubit oscillation period
$\tau_{meas}\gg 2\pi/\Omega$.  During the measurement, a well-defined
averaged current $(I_{23}+I_{14})/2$ appears.
Therefore, one can distinguish three subspaces
leading to effective measurement of the operator
$\sigma_y^{(1)} \sigma_y^{(2)} +\sigma_z^{(1)} \sigma_z^{(2)}$.
Contrary to naive expectation, weak
measurements taken for a long period of time are not equivalent to a
strong measurement because of non-trivial interplay between the qubit
Hamiltonian dynamics and dynamics due to the measurement process.

\subsection{Correlator of measurement current and power spectrum when $\epsilon = 0$ and $U=0$}

In this subsection we derive the current correlation function and
power spectrum of the detector signal for weak measurement of two
identical qubits symmetrically coupled to the detector.
It follows from (\ref{zee}) and
(\ref{noisy-curr-z})
that after collapse to a one-dimensional
subspace (to $|1\rangle^B$ in the linear and non-linear detector cases
and to $|1\rangle^B$ or $|2\rangle^B$ in the purely quadratic case)
the value of $z$ is constant, the qubits do not oscillate, and the
correlation function $K_I(\tau)=\langle I(t) I(t+\tau)\rangle -
\langle I \rangle^2$ contains only the noise. The corresponding power
spectrum is flat.  Collapse to a higher-dimensional subspace (the
three-dimensional spin-1 subspace in the linear detector case and the
two-dimensional
\{$|3\rangle^B,|4\rangle^B$\} subspace in the purely quadratic case)
leads to oscillations and a richer power spectrum that we derive below.

\subsubsection{Linear detection}

In the case of collapse to the spin-1 subspace in the linear case,
we have $\langle I \rangle = I_{23}$ and from Eq.\,(\ref{zee}),
$z = \rho_{11} - \rho_{44} = 2 \mbox{Re}\rho_{24}^B$.
Using expression
(\ref{noisy-curr-z}), we have
\begin{equation}
K_I(\tau >0)=(\Delta I)^2 K_z(\tau )+ \Delta I\, K_{\xi z}(\tau)
\label{KI}.
\end{equation}
First, we calculate the ``back-action correlator''
$K_{\xi z}(\tau )=\langle \xi(t) z(t+\tau)\rangle_t$ for $\tau>0$.
In an individual realization, the quantum back action on the qubit state $\rho_{ij}$
is taken into account explicitly through the Bayesian equations (\ref{Bayes1}).
In particular, $\rho_{ij}(t+\tau)$ is correlated with noise fluctuations of the current
at some previous moment $t$ (compare with Ref.\ \onlinecite{Mabuchi}).
Indeed, a noise fluctuation $\xi(t)$ leads to a
perturbation of the density matrix $\rho_{ij}(t)$ in the immediate
vicinity of the time instant $t$, i.e.,
$\rho_{ij}(t+0)=\rho_{ij}(t)+d\rho_{ij}(t)$.
The leading perturbation is given by Eq.(\ref{Bayesevolxi})
\begin{eqnarray}
dz(t) \simeq -2 \frac{\Delta I}{S_0} \left[ z^2(t)-(\rho_{22}^B(t)+\rho_{44}^B(t)) \right]\,
\xi(t)\,dt &&
\label{zxi-pert2}\\
dy(t) \simeq -2 \frac{\Delta I}{S_0} \left[ y(t) z(t) + y_1(t)/2 \right]\,
\xi(t)\,dt \qquad\quad &&
\label{yxi-pert2}
\end{eqnarray}
where $y = 2\mbox{Im}\rho_{23}^B$ and $y_1\equiv 2\mbox{Im}\rho_{34}^B$.

We proceed further by implicitly assuming that the long time average
$\langle\ldots\rangle_t$ is represented as three subsequent averages.
First, averaging over different realizations of the stochastic process
$\rho_{ij}(t')$ in the interval $t < t' < t+\tau$
is equivalent, assuming stationarity and ergodicity, to using the master
equations for $z(t)$ and $y(t)$
\begin{equation}
\dot{z} = -\Omega y, \,\,\,\, \dot{y} = \Omega z -\Gamma y.
\label{master-zy2}
\end{equation}
The ensemble dephasing rate is $\Gamma =\eta^{-1}(\Delta I)^2/4S_0$.
Note that the solution to the master equations is linear with respect
to initial conditions.
Explicitly, for $z(t)$ we have:
\begin{eqnarray}
&&z(t)=z(0)\,\exp (-\Gamma \tau /2)
[\cos \tilde\Omega \tau +(\Gamma /2\tilde\Omega)\sin \tilde\Omega \tau]
\nonumber\\
&&\qquad{}-y(0)\,\exp(-\Gamma \tau/2)
[\Omega /{\tilde\Omega}]\, \sin \tilde\Omega \tau
\nonumber\\
&&\qquad\equiv z(0)\,G_z(\tau) + y(0)\,G_y(\tau)
\label{Green-f} ,
\end{eqnarray}
where $\tilde\Omega \equiv (\Omega^2-\Gamma^2/4)^{1/2}$, and $G_z(\tau)$,
$G_y(\tau)$ are the corresponding Green's functions.
Therefore, the initial perturbation that appears due to $\xi(t)$,
propagates to the later moment $t+\tau$ according to  Eq.\,(\ref{Green-f})
with ``initial'' conditions $dz(t)$ and $dy(t)$:
$\delta z(t+\tau) = dz(t)\, G_z(\tau) + dy(t)\, G_y(\tau)$.
The second averaging is the standard averaging over noise  (at instant $t$) according to
$\langle \xi(t) \xi(t) dt\rangle = S_0/2$. The quantum correlation is expressed then as:
\begin{eqnarray}
\lefteqn{\langle \xi(t)\,z(t+\tau) \rangle =
\langle \xi(t)\,\delta z(t+\tau) \rangle }
\qquad\nonumber\\
&&{}=\Delta I
\left[\, \langle \rho_{22}^B(t)+\rho_{44}^B(t) - z^2(t) \rangle \,G_z(\tau) \right.
\qquad\qquad\quad\nonumber\\
&&\qquad\qquad\qquad\left. {}+\langle\, y(t)\,z(t) + y_1(t)/2\,\rangle \,G_y(\tau)\,\right].
\label{z-xi-corr2}
\end{eqnarray}
A third averaging over the ``initial'' condition $\rho_{ij}^B(t)$ remains to be done.

The calculation for the correlation $K_z(\tau) = \langle z(t) z(t+\tau)\rangle_t$
also proceeds in stages, first
averaging
over different realizations for $\rho_{ij}(t')$ in the interval $t < t' < t+\tau$.
Then
\begin{equation}
\langle z(t)\,z(t+\tau) \rangle = \langle z^2(t) \rangle \,G_z(\tau)
-\langle y(t)\,z(t) \rangle \,G_y(\tau)
\label{z-z-corr2} .
\end{equation}
In (\ref{KI}) the unknown averages
$\langle z^2(t) \rangle$ and $\langle y(t)\,z(t) \rangle$ cancel out,
and $\langle y_1 \rangle = 0$, so we obtain
\begin{equation}
K_I(\tau) = (\Delta I)^2 \langle\, \rho_{22}^B + \rho_{44}^B\,\rangle \,G_z(\tau)
= (\Delta I)^2 \frac{2}{3} \,G_z(\tau)
\label{I-I-corr2} ,
\end{equation}
where we have used stationary values from (\ref{stationary-lin}).
This is valid for arbitrary coupling ${\cal C}$ and detector ideality
$\eta$.

The corresponding current spectral density
$S_I(\omega) = 2\int_{-\infty}^{\infty} K_I(\tau)\exp(i\omega \tau) d\tau$
is readily computed by Fourier transform \cite{RusKor-ent} to be
\begin{equation}
    S_I(\omega ) = S_0 + \frac{8}{3}\, \frac{\Omega^2(\Delta I)^2\Gamma}
{(\omega^2-\Omega^2)^2+\Gamma^2\omega^2}.
\label{spectrum-lin}
\end{equation}
In the case of weak coupling, $\Gamma \ll \Omega$, the spectral peak is at
$\Omega$; it has the same width as in the one-qubit case \cite{Kor-osc}
and peak height $(32/3)\eta S_0$, which
we have confirmed in numerical simulations (Fig.\,\ref{fig-spectr}(a)).
It is important to emphasize that in an ensemble averaged approach
starting from a totally mixed state
this power spectrum
would be weighted by $3/4$, the probability of collapse
into the spin-1 subspace.  The averaged peak height would therefore be $8 \eta S_0$.
To derive the higher value of $(32/3) \eta S_0$ one must analyze a two-qubit system
already collapsed into the spin-1 subspace.

\subsubsection{Purely quadratic detection}

For purely quadratic detection, the system can collapse to $|1\rangle^B$ or to $|2\rangle^B$,
or it can collapse to
the subspace \{$|3\rangle^B,|4\rangle^B$\} and perform oscillations.  In this last scenario,
$\rho_{11}^B(t)=\rho_{22}^B(t)=0$ and
$\langle I \rangle= I_{23}-\Delta I/2$.
To study the oscillating signal, it is convenient
to rewrite Eq.\,
(\ref{noisy-curr-z}) in the form
$I(t)=\langle I \rangle + (\Delta I/2)\,z_1(t)+\xi(t)$ where $z_1(t)\equiv \rho_{33}^B(t) -\rho_{44}^B(t)$.
The It\^{o} equations for the
relevant density matrix components $z_1(t)$ and $y_1(t)$
again decouple from all other components
appearing in Eq.\,(\ref{Bayesevol})--(\ref{H-evol-U}).
The corresponding master equations read
\begin{equation}
\dot{z}_1 = -2\Omega y_1, \,\,\,\, \dot{y}_1 = 2\Omega z_1 -\Gamma y_1.
\end{equation}
Significantly,
motion in this two dimensional Hilbert space is precisely analogous to
one-qubit motion\cite{Kor-osc}, but with the oscillation frequency $2\Omega$ rather
than $\Omega$.  
This is natural because quadratic measurement is
sensitive only to the relative state of the two qubits
distinguishing only between states with ``parallel'' and ``anti-parallel'' spins.

Applying the same steps as in
Eqs. (\ref{zxi-pert2})-(\ref{z-z-corr2}), we obtain the power spectrum
\begin{equation}
S_I(\omega ) = S_0 + \frac{4\Omega^2(\Delta I)^2\Gamma}
{(\omega^2-4\Omega^2)^2+\Gamma^2\omega^2},
\label{spectrum-quadr}
\end{equation}
that exhibits a peak at $2\Omega$ with height $4\eta S_0$
and width $\Gamma  = (\Delta I)^2/4\eta S_0$ just
as in the linear case.\cite{Kor-osc,RusKor-sp,Bulaevskii,Shnirman-sp}
In an ensemble averaged approach,
starting from totally  mixed state
the peak height of
(\ref{spectrum-quadr}) would be weighted by $1/2$, the probability that
a totally mixed system will collapse into the subspace
\{$|3\rangle^B,|4\rangle^B$\}.  Thus the peak height would be $2\eta S_0$;
it is necessary to analyze a two-qubit system known to have collapsed
to the oscillating subspace
to predict the larger height of $4 \eta S_0$.

\subsection{ Influence of bias $\epsilon$ and two-qubit coupling $U$}
\label{twoqb-U}

We now discuss the influence of non-zero energy bias
$\varepsilon_a=\varepsilon_b=\varepsilon$ and two-qubit coupling $U$
for identical qubits.
Numerical simulations and analytical calculations both show that the
system still collapses and purifies.  However, differences arise in
the collapse scenarios.

In the case of a linear detector, the system will still collapse to
the spin-0 subspace $|1\rangle^B$ or to the spin-1 subspace
$\{|2\rangle^B,|3\rangle^B,|4\rangle^B\}$.
In fact, the corresponding stationary solution, Eq.\,(\ref{stationary-lin}),
remain unchanged.
This is because
the additional evolution due to $\varepsilon$- and $U$-terms,
Eqs. (\ref{H-evol-eps}) and (\ref{H-evol-U}), does not mix the two
subspaces.

The characteristics of the spin-0 collapse scenario are not
affected by the parameters $\epsilon$ and $U$.  In particular, the resulting power
spectrum is still flat with $S_I(\omega)$ equal to the detector shot noise
$S_0$.

In the oscillating, spin-1 scenario the additional dynamics do influence the
form of the spectral density $S_I(\omega)$.  For $U=0$ the relevant master equations are
\begin{eqnarray}
&& \dot{z} = -\Omega y, \,\,\,\,
\dot{y} = \Omega z -\Gamma y - (\varepsilon/\hbar) x,  \nonumber\\
&& \dot{x} = -\Gamma x + (\varepsilon/\hbar) y
\label{master-zyx}
\end{eqnarray}
with $x\equiv 2\mbox{Re}\rho_{34}^B$ and $y$, $z$ as in
Eq.\,(\ref{master-zy2}).
This is exactly the same
system of equations as in the one-qubit case with a non-zero bias.
Thus, the corresponding
spectral density in the two-qubit case is related to the
one-qubit spectral density peak,
$\tilde{S}_I^{1qb}(\omega,\Omega,\Gamma,\varepsilon) \equiv S_I^{1qb}-S_0$, considered in
Refs. \onlinecite{Kor-osc,RusKor-sp}
\begin{equation}
S_I^{2qb-linear}(\omega,\Omega,\Gamma,\varepsilon)=S_0 +
\frac{8}{3}\,\tilde{S}_I^{1qb}(\omega,\Omega,\Gamma,\varepsilon)
\label{relat-2-1} .
\end{equation}
As in the one-qubit case, finite $\varepsilon$ leads to a decrease of the spectral
peak around  the
Rabi frequency $\Omega$ and to the emergence of an extra peak around zero frequency.
Note, that the $8/3$ enhancement factor is just as in the
zero bias case, Eq.\,(\ref{spectrum-lin}).
Numerical simulations of
Eqs. (\ref{Bayesevol})--(\ref{H-evol-eps}) confirm this result.

For $\varepsilon = 0$ and $U\ne 0$,
the form of the spectral peak(s) in the oscillating scenario
can be calculated either by numerical
simulations using Bayesian Eqs. (\ref{Bayesevol})--(\ref{H-evol-U})
or through an ensemble average method developed in
Refs. \onlinecite{prl-quadr,Mao}.
Increasing $U$ from zero first suppresses and then splits
the spectral peak at the
Rabi frequency $\Omega$.

In the case of purely quadratic detection, qubit biasing so that $\varepsilon\ne 0$ while $U=0$
will lead to mixing between the $|2\rangle^B$ and \{$|3\rangle^B,|4\rangle^B$\}
subspaces as seen from Eq.\,(\ref{H-evol-eps}). For small bias
$\varepsilon$ this will take the form of rare switching events
whose rate and influence on the spectrum will be discussed in the next
section.

If $U\ne 0$ while $\varepsilon = 0$, one notices that the density matrix
components $z_1(t)\equiv \rho_{33}^B(t) -\rho_{44}^B(t)$,
$y_1(t)\equiv 2 \mbox{Im} \rho_{34}^B$, and $x\equiv 2\mbox{Re}\rho_{34}^B$
obey equations  analogous to (\ref{master-zyx}) in which
$\Omega$ is replaced by $2\Omega$, $z$ by $z_1$, $y$ by $y_1$, and
$\varepsilon$ by $2U$.
Thus, the form of the spectrum is described with the same functional form,
as in Eq.\,(\ref{relat-2-1})
\begin{equation}
S_I^{2qb-quadr}(\omega)=S_0 +
\tilde{S}_I^{1qb}(\omega,2\,\Omega,\Gamma,2\,U)
\label{relat-2-1-quadr}.
\end{equation}
The result is a decrease of the spectral peak at $2 \Omega$ seen
in Eq.\,(\ref{spectrum-quadr}) and the appearance of a peak at zero frequency.
This finding is confirmed by numerical simulations.\cite{prl-quadr}

\section{Quantum Zeno-like stabilization}
\label{stabilization}

The quantum Zeno effect is usually understood to arise in the
case of projective measurements
taken more and more frequently.  In our case of
continuous measurement the detector is always coupled to the system,
so the approach to the quantum Zeno regime corresponds instead to the
limit of stronger and stronger coupling.  Since it is impractical to
realize strong coupling detectors in many systems (the dimensionless
coupling
to each qubit, ${\cal C} \sim \Gamma/\Omega$, is typically small),
one might not expect quantum Zeno physics
to arise.  However, in our system of two identical qubits, a small
perturbative asymmetry between the qubits, such as when $\Delta
\Omega = \Omega_a - \Omega_b \neq 0$, can generate slow transitions between
states that were uncoupled.  The effective coupling for this dynamical
process, ${\cal C}_{eff} \sim \Gamma/\Delta \Omega$, can be very large,
leading to a Zeno effect.

From this perspective, the symmetric-qubit situation,
$\Delta \Omega = 0$, constitutes the Zeno limit of infinitely large coupling
${\cal C}_{eff}$; the transition becomes completely inactive, and the quantum
state becomes ``frozen''.  The asymmetric situation with non-zero
$\Delta \Omega$ implies finite effective coupling ${\cal C}_{eff}$.
The non-zero $\Delta \Omega$ makes the quantum state to ``decay'' from
one subspace to another. But, since the coupling is large, the decay
of the quantum state
looks like an abrupt {\it switching event}, with duration
on the order of the measurement time $\tau_{meas}\sim 1/\Gamma$,
rather than gradual evolution of frequency $\Delta \Omega$.  The switching event
can be identified experimentally
by measuring the power spectrum, for instance.
From an ensemble averaged point of view, at long times the decay
becomes exponential with a rate that is much smaller than $\Gamma$.

\begin{figure}
\centering
\includegraphics[width=3.2in]{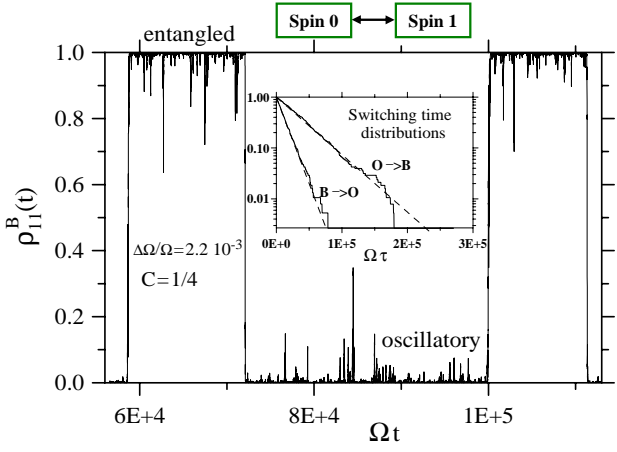}
\vspace*{0.25cm}

\caption{{\it Linear detection.}
Switching between spin-0 and spin-1 subspaces
as seen in the time evolution of $\rho_{11}^B(t)$,
in case of different Rabi frequencies.
The parameters are:
${\cal C}=1/4$, and $\Delta\Omega/\Omega\simeq 2.2\times 10^{-3}$.
The inset shows numerical histograms (in logarithmic scale)
of switching time distributions in agreement with the
asymptotic switching rates Eqs.\,(\ref{G1}),(\ref{reverse}).
}
\label{switching}
\end{figure}

Numerical simulations using
Bayesian equations (\ref{Bayes1}) confirm
that the two-qubit density matrix $\rho$ experiences
rare, abrupt switching between different subspaces when the qubits have a slight
asymmetry like a non-zero $\Delta \Omega$.  These switchings are
apparent in Figs.\ \ref{switching} and \ref{switch-eps}.  Switching also
happens for identical qubits when there is small difference of
couplings ${\cal C}_a$, ${\cal C}_b$ to the detector (due to different
values for $\Delta I_a$ and $\Delta I_b$) or different coupling to an
environment.

For linear detection
the switching occurs between the spin-0 and spin-1
subspaces  while for quadratic measurement we will have three-terminal
switching among the three collapse subspaces described above.
The switching events can be directly observed through
measurement of the {\it intermediate time scale} power spectrum and/or
average detector currents.
(Such intermediate time observables are meaningful if the measurement time
needed to accumulate the spectrum
is much smaller than the ``lifetime'' of the subspaces.)
If measurement is performed for a sufficiently long time, it will lead to
averaging of the spectrum and may also lead to the appearance of an additional
telegraph noise peak at zero frequency analogous to the regime of
strong coupling in the one-qubit case.\cite{Kor-osc,Makhlin}

\subsection{Perturbation to the qubits}

To find the switching rates analytically,
we have used the master equation, which reads
of Eqs. (\ref{Bayesevol})--(\ref{H-evol-eps})
$$\frac{d \rho_{ij}^B}{dt} = \left.\frac{d \rho_{ij}^B}{dt}\right|_{meas}  +
\left.\frac{d \rho_{ij}^B}{dt}\right|_{\Omega} + \left.\frac{d \rho_{ij}^B}{dt}\right|_{\varepsilon}
+ \left.\frac{d \rho_{ij}^B}{dt}\right|_{U}$$
with the noise term (\ref{Bayesevolxi}) omitted from
$\left.d \rho_{ij}^B/dt\right|_{meas}$.
It now acquires additional terms.
When the frequencies of the two qubits are slightly different
while coupling to the detector is symmetric, $I_2=I_3$,
we have in addition to Eq.\,(\ref{H-evol-om})
\begin{eqnarray}
\left.\frac{d \rho_{ij}^B}{dt}\right|_{H-evol,H_a\neq H_b}=\frac{\Delta\Omega}{2}\times
\qquad\qquad\qquad\qquad\qquad && \label{H-evol-mix1} \\
\left(
\begin{array}{c@{\:}c@{\:}c@{\:}c}
-2\,\mbox{Im}\rho_{12}^B\ & \imat (\rho_{11}^B-\rho_{22}^B)\ &-i \rho_{23}^B\ &-i \rho_{24}^B \\
\cdot                   \ & 2\,\mbox{Im}\rho_{12}^B        \ &-i \rho_{13}^B\ &-i \rho_{14}^B\\
\cdot                   \ & \cdot                         \ & 0            \ & 0           \\
 \cdot                  \ & \cdot                         \ & \cdot        \ & 0
\end{array}
\right) && \nonumber
\end{eqnarray}
and in addition to Eq.\,(\ref{H-evol-eps})
\begin{eqnarray}
\lefteqn{\left.\frac{d \rho_{ij}^B}{dt}\right|_{H-evol,\varepsilon_a \neq \varepsilon_b}=
- \frac{(\Delta\varepsilon/\hbar)}{2}\times
\qquad\qquad\qquad\qquad\qquad\qquad\qquad\qquad\qquad  } &&   \label{H-evol-mix2} \\
&& \left(\!\!
\begin{array}{c@{\:}c@{\:}c@{\:}c}
2\,\mbox{Im}\rho_{13}^B\ & i \rho_{32}^B\ &-i (\rho_{11}^B-\rho_{33}^B) & i \rho_{34}^B \\
\cdot                  \ & 0            \ &-i \rho_{21}^B           & -i (\rho_{22}^B-\rho_{44}^B)\\
\cdot                  \ & \cdot        \ &-2\,\mbox{Im}\rho_{13}^B & i \rho_{14}^B     \\
\cdot                  \ & \cdot        \ & \cdot                   & 0
\end{array}
\!\! \right)  \nonumber .
\end{eqnarray}

\subsubsection{Linear detection}

    For linear detection ($\delta=0$), assuming weak coupling
$\Delta \Omega \ll \Gamma \ll \Omega$ and
unbiased, non-interacting qubits $\varepsilon_a = \varepsilon_b=U=0$,
a system of equations for the quantities $\rho_{11}^B-\rho_{22}^B$,
$2\,\mbox{Im}\rho_{12}^B$, $\rho_{22}^B-\rho_{44}^B$,
$\rho_{33}^B-\rho_{44}^B$, and $2\,\mbox{Im}\rho_{34}^B$ can be
extracted from Eqs.\,(\ref{Bayesevol})--(\ref{H-evol-eps}) and
(\ref{H-evol-mix1}).
To change them into a system of master equations for ensemble averages,
we simply set the noise term to zero $\xi=0$ in (\ref{Bayesevol}).
Starting from the entangled initial condition $\rho_{11}^B(0)=1$
we  find
$\rho_{11}^B(t)=1-\frac{(\Delta\Omega)}{2}\int_0^t dt' \,2\,\mbox{Im}\rho_{12}^B(t')$.
Formal integration of the equation for
$2\,\mbox{Im}\rho_{12}^B$ (see Eqs. (\ref{Bayesevol}),
(\ref{H-evol-mix1})) leads to the exact relation
\begin{eqnarray}
\lefteqn{\rho_{11}^B(t)=1-\frac{(\Delta\Omega)^2}{2}\int_0^t dt' e^{-\Gamma t'} \times }
\qquad\qquad && \nonumber\\
&& \int_0^{t'} dt'' e^{\Gamma t''} \, (\rho_{11}^B(t'')-\rho_{22}^B(t''))
 \label{r11b-decay} .
\end{eqnarray}
Solving our system of master equations
to zeroth order in $\Delta\Omega$
and substituting into (\ref{r11b-decay}), we obtain an approximate solution for
the ``survival probability,'' $\rho_{11}^B(t)$:
\begin{equation}
\rho_{11}^B(t)=1+\frac{(\Delta\Omega)^2}{2\Gamma^2}-\frac{(\Delta\Omega)^2}{2\Gamma}\, t
-\frac{(\Delta\Omega)^2}{2\Gamma^2}\, \exp(-\Gamma t)
\label{approx-r11b}  .
\end{equation}
For times $t\ll 1/\Gamma$, one can expand the exponential and find that
there is no linear term in the  $t$-expansion of (\ref{approx-r11b}).
This fact makes quantum Zeno effect physics possible.\cite{Khalfin,MisraSudarshan,Peres}
The first few terms in the small $t$-expansion of (\ref{approx-r11b}) actually coincide with
the exact small $t$-expansion of $\rho_{11}^B(t)$
derived directly from the master equations
by  differentiation, using the initial conditions
\begin{equation}
\rho_{11}^B(t)=1-\frac{(\Delta\Omega)^2}{4}\, t^2+\frac{(\Delta\Omega)^2 \Gamma}{12}\, t^3
-\ldots
\label{small-t-r11b} .
\end{equation}
The $t^2$-coefficient turns out to be
$-(\Delta\Omega)^2/4 = -\mbox{Tr}[\hat{\rho}^2 {\cal H}^2_{QB}] +
\mbox{Tr}[\hat{\rho}{\cal H}_{QB}\hat{\rho}{\cal H}_{QB}]$,
i.e. it is determined by coherent (Hamiltonian) evolution alone
consistent with the discussions in Refs. \onlinecite{Khalfin,MisraSudarshan,Peres}.
One might speculate
that the system has some memory at the time scale $t \ll 1/\Gamma$,
``forgetting'' its history once $t\gtrsim \Gamma^{-1}$
when the exponential term in Eq.\,(\ref{approx-r11b}) drops out.
At still longer times, the ``decay'' from
the Bell state to the oscillating subspace eventually becomes exponential with rate
\begin{equation}
\Gamma_{B\rightarrow O}^{H_a\neq H_b}=\frac{(\Delta \Omega)^2}{2\Gamma}
\label{G1}
\end{equation}
as demonstrated by switching time distributions shown at the
inset of Fig.\,\ref{switching}.
Note, that the switching rate in (\ref{G1}) is similar to
the one-qubit case \cite{Kor-osc}
with the replacement $\Omega \rightarrow \Delta\Omega$.

    An analogous derivation for the case of different qubit biases
$\varepsilon_a \ne \varepsilon_b$ can be performed.
We set $\varepsilon=(\varepsilon_a+\varepsilon_b)/2=0$ since effects
of having $\varepsilon\ne 0$ (while $\varepsilon_a = \varepsilon_b$)
have been considered above, in Sec.\ref{twoqb-U}.
    We stipulate that $|\Delta\varepsilon|/\hbar \ll \Omega$
but permit $|\Delta\varepsilon|/\hbar$ and $\Gamma$ to be of the same order.
From (\ref{H-evol-om}), (\ref{H-evol-mix2}) we find
\begin{equation}
\dot{\rho}_{11}^B(t)=-(\Delta\varepsilon/\hbar)\,\mbox{Im}\rho_{13}^B = \dot{x}_{13}(t)/2
\label{r11b-decay-deps} ,
\end{equation}
where $x_{13}(t)\equiv \rho_{11}^B(t)-\rho_{33}^B(t)$.  The
second  equality in (\ref{r11b-decay-deps}) was obtained by
solving a system of nine equations extracted from
Eqs. (\ref{Bayesevol}), (\ref{H-evol-om}), and (\ref{H-evol-mix2}),
using the initial condition $\rho_{11}^B(0)=1$.
These nine equations quickly reduce to a system of three equations
for $2\,\mbox{Im}\rho_{13}^B$, $2\,\mbox{Re}\rho_{14}^B$, and
$x_{13}\equiv \rho_{11}^B-\rho_{33}^B$, which can be solved
perturbatively in $\Delta\varepsilon/\hbar$.
Integrating (\ref{r11b-decay-deps}), we obtain $\rho_{11}^B(t)=(1+x_{13}(t))/2$.
An approximate solution for $\rho_{11}^B(t)$ is
\begin{eqnarray}
\lefteqn{\rho_{11}^B(t)=1-b_{\varepsilon}\frac{\Gamma}{2}t+\frac{b_{\varepsilon} c_{\varepsilon}}{2}
-\frac{b_{\varepsilon}}{2}\, \exp(-\Gamma t/2)\times }\qquad\qquad\quad&& \nonumber\\
&& \left[
c_{\varepsilon} \cos{\widetilde{\Omega}t}-\frac{\Gamma}{2\widetilde{\Omega} }[2-c_{\varepsilon}]\sin{\widetilde{\Omega}t}\right]
\label{approx-r11b-deps}  ,
\end{eqnarray}
where $b_{\varepsilon}\equiv
(\Delta\varepsilon/\hbar)^2/[\Omega^2+(\Delta\varepsilon/\hbar)^2]$,
$c_{\varepsilon}\equiv \Gamma^2/[\Omega^2+(\Delta\varepsilon/\hbar)^2]
-1$, and $\widetilde{\Omega}\equiv
\sqrt{\Omega^2+(\Delta\varepsilon/\hbar)^2-\Gamma^2/4}$. As in
(\ref{small-t-r11b}), there is no linear term; the evolution is
quadratic in $t$ at small times permitting quantum Zeno effect
physics.\cite{Peres} The decay is exponential on a coarse time scale $t\gtrsim
2 \Gamma^{-1}$ with switching rate
\begin{equation}
\Gamma_{B\rightarrow O}^{\varepsilon_a\neq \varepsilon_b}=
\frac{(\Delta \varepsilon/\hbar)^2\ \Gamma}{2\Omega^2}
\label{G4}.
\end{equation}
Numerical solution of the master equations for the density matrix
confirms these results.

We have computed the rate of switching from the spin-0 Bell state $|1\rangle^B$ to the spin-1 subspace.
To compute the rate of reverse switching, we notice
that the stationary solution of the master equations
(Eqs. (\ref{Bayesevol})--(\ref{H-evol-U}), (\ref{H-evol-mix1}), and (\ref{H-evol-mix2})
with the noise $\xi$ set to zero) has the form
\begin{equation}
\rho_{ij,st}^B=\rho_{ij,st}=\delta_{ij}/4.
\label{mix}
\end{equation}
This form is completely mixed because (\ref{H-evol-mix1}) and (\ref{H-evol-mix2})
cause transitions between the spin-0 and spin-1 subspaces.
One infers that the system should spend on average $1/4$ of the time in
the state $|1\rangle^B$, suggesting the stationary state condition
$(1/4)\Gamma_{B\rightarrow O} - (3/4)\Gamma_{O\rightarrow B} = 0$ or
\begin{equation}
\Gamma_{O\rightarrow B} = \Gamma_{B\rightarrow O}/3.
\label{reverse}
\end{equation}
The numerical histograms of switching time distributions
(see inset of Fig.\, \ref{switching}), which were
accumulated in a Monte-Carlo computation, confirm
the long-time exponential decay of the ``survival'' probability
$\rho^B_{11}(t)$ as well as the asymptotic rates (\ref{G1}),(\ref{G4}), and (\ref{reverse}).

If a state's lifetime in the oscillatory subspace,
$\tau_O\equiv 1/\Gamma_{O\rightarrow B}$, is sufficiently long, it is
possible to measure the spectrum described by Eq.\,(\ref{spectrum-lin}).
The necessary condition for this is
$\tau_O \gtrsim \tau_{meas}$
where $\tau_{meas} \sim 1/\Gamma$ is a typical measurement time needed to achieve
signal-to-noise ratio close to unity. \cite{Kor-99-01}
For small efficiency $\eta$ we estimate (see Appendix B)
\begin{equation}
\tau_{meas}\simeq \frac{9}{32\,\Gamma \eta^2}
\label{time-meas}
\end{equation}
while for $\eta$ close to unity the measurement time may become
an order of magnitude larger than (\ref{time-meas}).
Physically, the lifetime will satisfy $\tau_O \gtrsim \tau_{meas}$ if one can
tune close to the symmetric point of identical qubits (small $\Delta \Omega$)
and if the efficiency $\eta$ is not too small.

The spectrum accumulated for times much
greater than the lifetime $\tau_O$ gets averaged by switching events.
Since the
system resides in the spin-0 subspace $1/4$ of the time and in the spin-1
subspace the remaining $3/4$, the spectrum will take the form
(\ref{spectrum-lin}) with a factor of $2$
instead of $8/3$
(i.e. the peak-to-pedestal ratio will be $8 \eta$).
There will be no additional peak at zero frequency because the
averaged currents in the two scenarios coincide.

\subsubsection{Purely quadratic detection}

For a purely quadratic detector ($\delta=-2$) in the case of
$\Gamma \gg \Delta\Omega\neq 0$, Eq.\,(\ref{H-evol-mix1}) produces
switching transitions between states  $|1\rangle^B$ and $|2\rangle^B$
while the oscillating state inside $\{|3\rangle^B,|4\rangle^B\}$ does not switch.
The stationary solution to the master equation is similar to (\ref{stationary-quadr}),
but now $\rho_{11,st}^B=\rho_{22,st}^B=\frac{1}{2}[\rho_{11}^B(0)+\rho_{22}^B(0)]$.
The new measurement subspaces are the minus subspace
``$-$''$=\{|1\rangle^B,|2\rangle^B\}$ and the plus subspace
``$+$''$=\{|3\rangle^B,|4\rangle^B\}$, where the names are based
upon (\ref{Bell1}) - (\ref{Bell4}).
The switching rates are given by
\begin{equation}
\Gamma_{|1\rangle^B\to|2\rangle^B}=\Gamma_{|2\rangle^B\to|1\rangle^B}=
\frac{(\Delta \Omega)^2}{2\Gamma}
\label{G1-a} ,
\end{equation}
derived by solving master equations to lowest order in $\Delta \Omega$ as
in the derivation of Eq.\,(\ref{G1}).
Switching between $|1\rangle^B$ and $|2\rangle^B$,
and therefore between currents $I_{23}$ and $I_{14}$ in Fig.\,\ref{levels},
produces telegraph noise.\cite{Machlup,Kogan}  As a result, if we measure long enough
($t\gg 1/\Gamma_{|1\rangle^B\to|2\rangle^B}$), the
power spectrum will exhibit a peak at zero frequency
\begin{equation}
S_{I}(\omega )= S_0 + \frac{(\Delta I)^2 \Gamma}{(\Delta \Omega)^2}
\, \frac{1}{1+[\omega \Gamma /(\Delta \Omega)^2]^2}
\label{small-deomega}
\end{equation}
in full analogy with the one-qubit case.

Actually, it is possible to derive the power spectrum for arbitrarily large $\Delta \Omega$:
\begin{equation}
S_I(\omega ) = S_0 + \frac{(\Delta\Omega)^2(\Delta I)^2\Gamma}
{[\omega^2-(\Delta\Omega)^2]^2+\Gamma^2\omega^2}
\label{arbitr-deomega}
\end{equation}
which reduces to (\ref{small-deomega}) for small $\Delta \Omega$.
Indeed, after collapse to the ``$-$'' or ``$+$'' subspaces,
Eqs. (\ref{free-evol12}), (\ref{free-evol34}), (\ref{Bayesevol}), and (\ref{Bayesevolxi})
produce weakly perturbed
quantum oscillations within the subspace and no switching to the other subspace.
The quantum oscillations in the minus and plus subspaces
correspond directly to
one-qubit dynamics with oscillation frequency $\Delta\Omega$ and
frequency $2\Omega \equiv (\Omega_a + \Omega_b)$ respectively.
After a collapse in the ``$-$'' subspace the spectrum will be given by
(\ref{arbitr-deomega}).
If collapse has happened in the ``$+$'' subspace the spectrum will be
given by Eq.\,(\ref{spectrum-quadr}) with a peak at $2\Omega$.
In both cases the maximum peak-to-pedestal ratio at a non-zero frequency is $4 \eta$.

Measuring the spectrum on a sufficiently long time scale will
constitute an (effective) measurement of the operator $\sigma_x^{(1)}
\sigma_x^{(2)}$ whose eigenspaces are the ``$-$'' and ``$+$''
subspaces.

\begin{figure}
\centering
\includegraphics[width=3.05in]{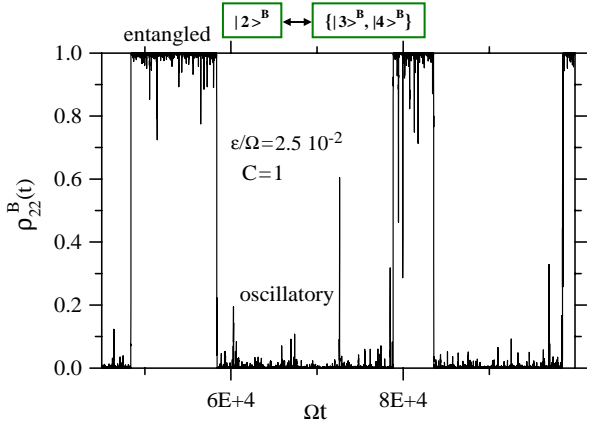}
\vspace{0.4cm}
\caption{{\it Quadratic detection.}
Switching between Bell subspaces
$|2\rangle^B$ and $\{|3\rangle^B,|4\rangle^B\}$ of Eqs.\, (\ref{Bell1}) -- (\ref{Bell4})
as seen in the time evolution of $\rho_{22}^B(t)$,
due to a small qubit bias. $|1\rangle^B$ is unaffected as seen from (\ref{H-evol-eps}).
The parameters are:
${\cal C}=1$, and $\varepsilon/\Omega\simeq 2.5\times 10^{-2}$.
}
\label{switch-eps}
\end{figure}

As we mentioned in Sec.\ref{twoqb-U}, if instead of an asymmetry in frequency,
the qubits are identical and biased $\varepsilon_a=\varepsilon_b=\varepsilon \ll \Gamma \ll \Omega$,
under quadratic measurement switching will occur between the $|2\rangle^B$ state
and the $\{|3\rangle^B,|4\rangle^B\}$ subspace (see Eq.\,(\ref{H-evol-eps})).
The spin-0 subspace $|1\rangle^B$ will be unaffected as shown in Fig.\,\ref{switch-eps}.

The asymptotic switching rates (relevant at times $t \gtrsim 1/\Gamma$)
are derived by study of master equations.  The stationary solutions
are given by (\ref{stationary-lin}) and the switching rate by
\begin{equation}
\Gamma_{|2\rangle^B\to\{|3\rangle^B,|4\rangle^B\}}=
\frac{2\varepsilon^2}{\Omega^2} \Gamma
        \label{G-eps0}.
\end{equation}
Since $\rho_{22,st}^B=\rho_{33,st}^B=\rho_{44,st}^B=1/3$, $\rho_{11,st}^B = \rho_{i\neq j,st}^B=0$
is a stationary solution of the master equation
within the spin-1 subspace, the stationary condition must be
$(1/3) \Gamma_{|2\rangle^B\to\{|3\rangle^B,|4\rangle^B\}} - (2/3) \Gamma_{\{|3\rangle^B,|4\rangle^B\}\to|2\rangle^B}= 0$,
implying
\begin{equation}
\Gamma_{\{|3\rangle^B,|4\rangle^B\}\to|2\rangle^B}=
\frac{1}{2}\, \Gamma_{|2\rangle^B\to\{|3\rangle^B,|4\rangle^B\}}
\label{G-eps}.
\end{equation}

When the system collapses to $|1\rangle^B$, the power spectrum remains flat
$S_I(\omega) = S_0$. In the case of switching between $|2\rangle^B$ and
$\{|3\rangle^B,|4\rangle^B\}$, we can average the spectra
as in the linear case above
and add a telegraph noise peak
caused by switching between current levels $I_{14}$ and $(I_{23}+I_{14})/2$.
The result is
\begin{eqnarray}
&& S_{I}(\omega )= S_0 +
\frac{2}{3} \frac{4\Omega^2(\Delta I)^2\Gamma}
{(\omega^2-4\Omega^2)^2+\Gamma^2\omega^2}
 \nonumber \\
&& \hspace{1cm} +
\frac{2 (\Delta I)^2 \Omega^2}{27 \varepsilon^2 \Gamma}
\, \frac{1}{1+[ \omega  \Omega^2/3 \varepsilon^2 \Gamma ]^2}
        \label{S_eps}.
        \end{eqnarray}
Note that the peak-to-pedestal ratio at $2\Omega$ is multiplied by a factor $2/3$
as a result of the averaging.  This result can be reproduced
also in the ensemble averaged approach.\cite{prl-quadr,Mao}

To conclude this section we mention that a non-zero qubit interaction $U$
does not mix the measurement subspaces, neither in the linear nor in the quadratic case.
Correspondingly, calculations show that neither the stationary solutions
of the master equations nor
the switching rates derived above will be affected.
Finite $U$ will only affect
the form of the spectrum  in the oscillating scenario
as described in Sec.\ref{twoqb-U}.

\subsection{Slightly different couplings}

\begin{figure}
\centering
\includegraphics[width=3.in]{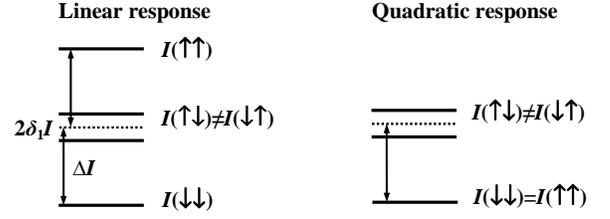}
\vspace*{0.2cm}

\caption{Current levels for asymmetrically coupled qubits,
$\Delta I_a \neq \Delta I_b$.
Linear and quadratic response.}
\label{levels-asym}
\end{figure}

For identical qubits with slightly different coupling to the detector,
$I_2\neq I_3$ as indicated in Fig.\ \ref{levels-asym}.  This asymmetry
introduces evolution between collapse subspaces.  The presence of the
detector provides quantum Zeno stabilization, and so instead of
gradual evolution, there is rare, abrupt switching between the
subspaces.  It is convenient to introduce the current asymmetry
$\delta_1 I\equiv (I_2-I_3)/2=(\Delta I_a - \Delta I_b)/2$ and to
define the current difference as $\Delta I = \Delta I_a - \delta_1 I =
\Delta I_b + \delta_1 I=(\Delta I_a + \Delta I_b)/2$.  The current
asymmetry parameter is $\delta_1 \equiv \delta_1 I/\Delta I$.
Defining the difference in coupling by $\Delta {\cal C} = {\cal C}_a - {\cal C}_b$
and the average coupling by ${\cal C} = ({\cal C}_a + {\cal C}_b)/2$,
we note that in cases of small coupling asymmetry,
$\delta_1 \simeq \frac{\Delta {\cal C}}{4\, {\cal C}}$.

\subsubsection{Linear detection}

If $\delta_1 \ne 0$, the following additional terms arise in
the master equation in the case of linear detection
\begin{eqnarray}
\lefteqn{\left.\frac{d \rho_{ij}^B}{dt}\right|_{meas,mix}\!\!\! =
 \frac{(\Delta I)^2}{\eta 4S_0}\,\left[ 2\delta_1
\left(
\begin{array}{cccc}
0      & \rho_{34}^B& 0           & \rho_{32}^B \\
\cdot  & 0          & \rho_{41}^B & 0           \\
\cdot  & \cdot      & 0           & \rho_{12}^B \\
\cdot  & \cdot      & \cdot       & 0
\end{array}
\right) \right. } && \!\!\! \label{Bayesevol-mix} \\
&& \left. -2 (\delta_1)^2 \left(
\begin{array}{cccc}
(\rho_{11}^B-\rho_{33}^B) & \rho_{12}^B/2 & 2i\mbox{Im}\rho_{13}^B    & \rho_{14}^B/2 \\
\cdot                     & 0             & \rho_{23}^B/2             & 0 \\
\cdot                     & \cdot         & (\rho_{33}^B-\rho_{11}^B) & \rho_{34}^B/2 \\
\cdot                     & \cdot         & \cdot                     & 0
\end{array}
\right) \right] \nonumber
\end{eqnarray}
If the system starts in the entangled initial condition, $\rho_{11}^B=1$,
the exact solution to the master equation for small times is
\begin{equation}
\rho_{11}^B(t)=1-2 (\Gamma \delta_1^2)\, t +4 (\Gamma \delta_1^2)^2\, t^2+\cdots
\label{rho11smallt}
\end{equation}
The solution contains a linear term setting a time scale $(\Gamma \delta_1 ^2)^{-1}\gg \Gamma^{-1}$.
For $\rho_{11}^B(t)$ close to unity, the asymptotic rate
of switching from the spin-0 Bell state $|1\rangle^B$ to the spin-1 subspace is found to be
\begin{equation}
\Gamma_{B\rightarrow O}^{{\cal C}_a\neq {\cal C}_b} = 2\Gamma \delta_1^2 =\frac{(I_2 - I_3)^2}{\eta 8 S_0}\cong
\frac{1}{8}\left(\frac{\Delta {\cal C}}{{\cal C}}\right)^2 \Gamma
\label{G2} ,
\end{equation}
where the qubit detector couplings are slightly different, $\Delta I_a \neq \Delta I_b$.
The reverse switching rate is given by Eq.\,(\ref{reverse}).

To understand Eq.\,(\ref{G2}), we note that measurement for a short
interval $\delta t$ cannot distinguish between the current levels
$I_2$ and $I_3$ when $(I_2 - I_3)^2 \ll D_I$ where $D_I=S_0/2\delta
t$ is the variance of the measured noisy current.  For a system
initially in the spin-0 state $|1\rangle^B \sim |2\rangle - |3\rangle$
weak measurement tends to collapse the system towards either
$|2\rangle$ or $|3\rangle$, thereby diminishing the amount of
entangled state $|1\rangle^B$.  Standard decoherence\cite{Davies} in
the measurement basis
$\rho_{23}\rightarrow \rho_{23} \exp(-(I_2 - I_3)^2/8 D_I)$ leads in the Bell basis to
$\rho_{11}^B(t)\simeq 1-[(I_2 - I_3)^2/8 S_0]\,\delta t$ in agreement with (\ref{G2}).
Note that the rate (\ref{G2}) is one half of the average classical information
acquisition rate\cite{Kor-ent,Clerk} ($\Gamma_{cl}=(I_2-I_3)^2/\eta 4S_0$)
for a weakly measured fictitious two level system, as expected.

\subsubsection{Quadratic detection}

In the case of quadratic detection, if $I_2$ and $I_3$ are slightly
different while $I_1$ remains equal to $I_4$, then the two qubits
have slightly different couplings and also the detector deviates
slightly from being purely quadratic.  The additional terms  in the
master equations take the form (\ref{Bayesevol-mix}) but
with the index ``$2$'' interchanged with the index ``$4$'' everywhere in the first
matrix.
As a result, the state $|2\rangle^B$ is
unaffected, while rare switching  develops between the state
$|1\rangle^B$ and oscillations in the subspace
$\{|3\rangle^B,|4\rangle^B\}$.
The small time expansion of $\rho_{11}^B(t)$ takes the same form (\ref{rho11smallt}) and thus
the rate $\Gamma_{|1\rangle^B\to\{|3\rangle^B,|4\rangle^B\}}$ is the
same as in the linear case, Eq.\ (\ref{G2}):
\begin{equation}
\Gamma_{|1\rangle^B\to\{|3\rangle^B,|4\rangle^B\}}=
\frac{\Gamma}{2} \left(\frac{I_2-I_3}{\Delta I}\right)^2 ,
\label{G23}
\end{equation}
Since $\rho_{11,st}^B=\rho_{33,st}^B=\rho_{44,st}^B=1/3$, $\rho_{i\neq j,st}^B=0$ is
a stationary solution to the master equation
within the subspace $\{|1\rangle^B,|3\rangle^B,|4\rangle^B\}$),
 the  reverse switching rate is
$\Gamma_{\{|3\rangle^B,|4\rangle^B\}\to|1\rangle^B}= \frac{1}{2}\, \Gamma_{|1\rangle^B\to\{|3\rangle^B,|4\rangle^B\}}$.
In the case of collapse to $|2\rangle^B$, the spectrum remains flat, $S_I(\omega) = S_0$.
Otherwise, for measurements over a time interval too short for switching, the system will
reside in $|1\rangle^B$  leading to $S_I(\omega) = S_0$ or in the subspace $\{|3\rangle^B,|4\rangle^B\}$,
leading to Eq.\ (\ref{spectrum-quadr}).  For longer times, the spectrum will be the average
of $S_I(\omega) = S_0$ and the spectrum given by
Eq.\,(\ref{spectrum-quadr})
together with a telegraph noise peak:
\begin{eqnarray}
\lefteqn{S_{I}(\omega )= S_0 +
\frac{2}{3} \frac{4\Omega^2(\Delta I)^2\Gamma}
{(\omega^2-4\Omega^2)^2+\Gamma^2\omega^2} } &&
 \nonumber \\
&& \!\!\!\!\!\!\! + \frac{8 (\Delta I)^4}{27\, \Gamma (I_2-I_3)^2}
\, \frac{1}{1+[4 \omega  (\Delta I)^2/3\Gamma (I_2-I_3)^2]^2}.
\label{S_23}
\end{eqnarray}
The peak-to-pedestal ratio at $2\Omega$ will be $(8/3)\eta$ rather than $4\eta$ as a result
of the averaging.

If the detector has $I_1\neq I_4$ while $I_2 = I_3$,
switching occurs between the state $|2\rangle^B$ and the oscillating subspace
$\{|3\rangle^B,|4\rangle^B\}$, while state $|1\rangle^B$ is unaffected
(as in Fig.\ref{switch-eps}). This slightly non-quadratic
mode of detection is ``orthogonal'' to the $I_2 \neq I_3$ mode
in the sense that the switching rate  would not depend on
a small additional $I_2 \neq I_3$- asymmetry.
The rates are
\begin{eqnarray}
&& \Gamma_{|2\rangle^B\to\{|3\rangle^B,|4\rangle^B\}}=
\frac{\Gamma}{2} \left(\frac{I_1-I_4}{\Delta I}\right)^2 ,
        \label{G14} \\
&&
\Gamma_{\{|3\rangle^B,|4\rangle^B\}\to|2\rangle^B}=
\frac{1}{2}\, \Gamma_{|2\rangle^B\to\{|3\rangle^B,|4\rangle^B\}}.
\label{G14r}
\end{eqnarray}
The power spectrum for large times will be given by Eq.\ (\ref{S_23})
with the substitution $(I_2-I_3)\rightarrow (I_1-I_4)$.

When $I_2\neq I_3$ {\it and} $I_1\neq I_4$, we combine the two cases
$I_2 \neq I_3$, $I_1=I_4$ and $I_2 = I_3$, $I_1\neq I_4$.  Based on
the analysis above, there should be switching $|1\rangle^B
\leftrightarrow \{|3\rangle^B,|4\rangle^B \} \leftrightarrow
|2\rangle^B$ but no direct switching between $|1\rangle^B$
and $|2\rangle^B$. The stationary solution of the master equation is
$\rho_{ii,st}^B=1/4$, $\rho_{i\neq j,st}^B=0$, so only one half of the
time is spent in the oscillatory state, and therefore the
peak-to-pedestal ratio at $2\Omega$ is $2\eta$ rather than $4\eta$.
The current spectral density in this case, including a
low-frequency peak due to telegraph noise in a 3-terminal system, is
\begin{eqnarray}
&& S_I(\omega) =  S_0 +
\frac{1}{2} \frac{4\Omega^2(\Delta I)^2\Gamma}
{(\omega^2-4\Omega^2)^2+\Gamma^2\omega^2}
    \nonumber \\
&& 
+ \frac{(\Delta I)^2 (\Gamma_1+\Gamma_2)(4\Gamma_1\Gamma_2+\omega^2)}
{16\Gamma_1^2\Gamma_2^2 + (9\Gamma_1^2 + 2\Gamma_1\Gamma_2 +
9\Gamma_2^2)\omega^2 +4\omega^4}
\label{3-telegraph}
\end{eqnarray}
where $\Gamma_1\equiv\Gamma_{|1\rangle^B\to\{|3\rangle^B,|4\rangle^B\}}$ and
$\Gamma_2\equiv\Gamma_{|2\rangle^B\to\{|3\rangle^B,|4\rangle^B\}}$
are given by (\ref{G23}) and (\ref{G14}).

\subsection{Environmental dephasing}

We have studied the effect of environmental dephasing,
modeling it with two small dephasing rates $\gamma_a$ and $\gamma_b$ acting
separately on the qubits.
This adds an extra dephasing contribution $\widetilde{\gamma}_{ij}$
to $\gamma_{ij}$ in Bayesian equation (\ref{Bayes1}) where
\begin{eqnarray}
\widetilde{\gamma}_{ij} =
\left(
\begin{array}{cccc}
0                & \gamma_b         & \gamma_a         & \gamma_a+\gamma_b \\
\gamma_b         & 0                & \gamma_a+\gamma_b& \gamma_a          \\
\gamma_a         & \gamma_a+\gamma_b& 0                & \gamma_b \\
\gamma_a+\gamma_b& \gamma_a         & \gamma_b         & 0
\end{array}
\right).  &&
\label{env-dephasing}
\end{eqnarray}
We assume that the coupling to the environment is much smaller than
the coupling to the detector, $\gamma_a,\gamma_b\ll\Gamma$.
Numerical simulation again shows that abrupt switching events take
place.  However, now the density matrix $\rho$ remains slightly mixed
even for an ideal detector.

In the linear case we have switching between spin-0 and spin-1
subspaces.  Analyzing the resulting master equations, we derive in the
case of a linear detector the switching rates
\begin{equation}
\Gamma_{B\rightarrow O}^{\gamma_a,\gamma_b}=
3\Gamma_{O\rightarrow B}^{\gamma_a,\gamma_b}=(\gamma_a+\gamma_b)/2.
\label{G3-lin}
\end{equation}
In the quadratic case the switching involves transitions $|1\rangle^B
\leftrightarrow \{|3\rangle^B,|4\rangle^B\} \leftrightarrow
|2\rangle^B$, with again no direct switching between $|1\rangle^B$ and
$|2\rangle^B$.  The rates are
$\Gamma_{|1\rangle^B\to\{|3\rangle^B,|4\rangle^B\}}=
\Gamma_{|2\rangle^B\to\{|3\rangle^B,|4\rangle^B\}}=(\gamma_a+\gamma_b)/2$,
and the reverse switching rates follow from the stationary state.

\section{Conclusion}

In this paper we have investigated in detail the non-trivial interplay
between the internal two-qubit Hamiltonian evolution and two-qubit
dynamics under weak continuous measurement described by a Bayesian
formalism.  The gradual collapse of the system's quantum state takes a
finite measurement time $\tau_{meas} \sim 1/\Gamma$.
On a time scale greater than $1/\Gamma$, the
measurement effectively looks like orthodox projective measurement in the Bell
basis.  A linear or non-linear detector measures the operator of total ``spin'',
$(\overrightarrow{\sigma}^{(1)} + \overrightarrow{\sigma}^{(2)})^2$.
The system's possible collapse
subspaces are the spin-0 Bell state, $|1\rangle^B$ and the spin-1
subspace within which the system executes oscillations.  For purely
quadratic measurement, the operator
$\sigma_y^{(1)} \sigma_y^{(2)} +\sigma_z^{(1)} \sigma_z^{(2)}$
is measured so that the collapse
subspaces are the spin-0 state, the other Bell state, $|2\rangle^B$,
and the oscillation subspace \{$|3\rangle^B$, $|4\rangle^B$\}.

Measurement leads to spontaneous entanglement which can be
identified by a distinct power spectrum and/or distinct value of the
average current of the detector.  We obtained analytically the power
spectrum of the current, $S_I(\omega)$, in different situations.  In
particular, in the case of a linear (purely quadratic) detector
response for non-interacting identical qubits the spectrum exhibits a
Lorentzian single peak at frequency $\Omega$ ($2\Omega$) with
peak-to-pedestal ratio of $(32/3)\eta$ ($4\eta$) respectively.  We
also obtained analytic expressions in the case of qubit biasing and in
the presence of qubit-qubit interaction. For a general non-linear
measurement the spectrum can be calculated using numerical
simulations of the measurement process.

Various small imperfections of the two-qubit system such as asymmetry
of the Hamiltonian and/or asymmetry in the coupling to detector cause
transitions between collapsed states.  Because the coupling to the
detector, although weak compared to most of the energy scales in the
Hamiltonian, can be strong compared to the small imperfection, a
regime of quantum Zeno-like stabilization arises.  The system does not
drift gradually from one collapse subspace to another; instead it
remains stabilized in one subspace and experiences rare, abrupt
switching events to other subspaces.  For very long observational times,
because of the switching, one observes only averaged dynamics in the power spectrum.

The authors gratefully acknowledge the support of the Packard foundation
(R.R. and A.M.) and of NSA and ARDA under ARO grant W911NF-04-1-0204 (A.K.).

\appendix

\section{ Calculation of the stationary distribution for $\rho_{22}^B$}

In this Appendix,
we use a Fokker-Plank equation to calculate the stationary distribution
for $\rho_{22}^B$ and its moments in the case of linear detection
in the oscillating scenario of Fig.\,\ref{fig-pur-osc} and
Eqs. (\ref{z-parameter})--(\ref{y1-parameter}).

An It\^o stochastic equation for
$\rho_{22}^B$ can be derived from Eqs. (\ref{Bayesevol})--(\ref{H-evol-om}), and the
parameterizations (\ref{z-parameter})--(\ref{y1-parameter}), that depends
only on the two independent variables, $x(t)\equiv \rho_{22}^B(t)$ and $\varphi(t)$:
\begin{eqnarray}
&&\frac{dx}{dt} = 2 \frac{\Delta I}{S_0} \sqrt{x(1-x)}\,(1-2x) \cos(\Omega t + \varphi(t))\ \xi(t)
\nonumber\\
&&\qquad\quad{}-2 \Gamma\, [x-(1-x) \cos^2(\Omega t + \varphi(t)) ]
\label{x-Ito} .
\end{eqnarray}
In the weak coupling regime one can average over the oscillations with
frequency $\Omega$. Then $\langle \cos^2(\Omega t + \varphi(t)) \rangle \to 1/2$ and
\begin{equation}
\frac{dx}{dt} = \frac{\Delta I}{S_0} \sqrt{2 x(1-x)}\,(1-2x)\ \xi_1(t)
-\Gamma\,(3x-1)
\label{x-Ito-average} ,
\end{equation}
where, $\xi_1(t)$ is an effective white noise with the same spectral density.

A Fokker-Plank equation for the distribution $P(x,t)$
follows\cite{Gardiner,Kogan} from Eq.\,(\ref{x-Ito-average})
\begin{equation}
\frac{\partial P(x,t)}{dt} = -\frac{\partial}{\partial x} [B(x) P(x,t)]
+ \frac{S_0}{4}\frac{\partial^2}{\partial x^2} [C^2(x) P(x,t)]
\label{FP}
\end{equation}
with $B(x)\equiv -\Gamma(3x-1)$ and $C(x)\equiv \frac{\Delta I}{S_0}\sqrt{2x(1-x)}$.
A stationary solution of Eq.\,(\ref{FP})   
can be derived\cite{Gardiner} taking into account that $x=\rho_{22}^B$ is a restricted
variable   
and therefore the ``probability current flow'' is zero.
\begin{equation}
P(x) = N_0^{-1}\
\frac{\exp{\displaystyle \left[ -\frac{1}{2(1-2x)}\right] }}{x^{1/2} (1-2x)^{7/2}}
\label{P(x)}
\end{equation}
with normalization constant $N_0 = 6 \sqrt{\pi/e}$.
Eq.\, (\ref{P(x)}) implies that $\rho_{22}^B$ actually fluctuates
between 0 and 1/2,
which is confirmed by 
Monte-Carlo simulations, Fig.\,\ref{fig-pur-osc}.
For the moments,
$\langle x^n \rangle\equiv \int_0^{1/2} x^n P(x)\, dx$, one obtains
\begin{eqnarray}
&& \langle x \rangle = 1/3, \qquad \langle x^2 \rangle = 1/8,\ \ \ \cdots \label{moment12}\\
&& \langle x^n \rangle = (N_0)^{-1}
\left[ 3\sqrt{2\pi}\,{}_{{}_{\scriptstyle 1}}F_{\scriptstyle 1}(\frac{1}{2}-n,-\frac{3}{2};-\frac{1}{2})
\right. \label{momentn}\\
&&\qquad\left.{}-\frac{8\sqrt{\pi}}{15} \frac{\Gamma(n+\frac{1}{2})}{\Gamma(n-2)}
\,{}_{{}_{\scriptstyle 1}}F_{\scriptstyle 1}(3-n,\frac{7}{2};-\frac{1}{2}) \right] 2^{1/2-n}
\nonumber
\end{eqnarray}
where ${}_{{}_{\scriptstyle 1}}F_{\scriptstyle 1}(a,b;z)$ is the confluent hypergeometric function.

\section{Noise of the noise and the measurement time to accumulate the spectrum}

Here we calculate the measurement time to distinguish between a flat spectrum
(shot noise) and a spectrum that possesses a peak on the top of the shot noise pedestal.
To make an estimation we first consider a situation with a narrow band
($\Delta f \ll f$, $f\equiv \omega/2\pi$) frequency filter.
The average power for a finite time $\Delta t$ ($\geq 1/\Delta f$)
is a fluctuating quantity,
${\cal P} = \frac{1}{\Delta t} \int_{-\frac{\Delta t}{2}}^{\frac{\Delta t}{2}}
    \,dt\, \left(I(t)-I_0 \right)^2$,
that depends on $\Delta t$ and on realizations of the process.
Denoting the  fluctuating part of the current as
${\tilde I}(t) \equiv I(t) - I_0$ we perform
averaging over realizations  
using the well known relation to the noise power spectrum
(see, e.g. Ref. \onlinecite{Kogan}):
$\langle {\tilde I}^2(t) \rangle_r \simeq S_I(\omega) \Delta f$.
Therefore, averaging of ${\cal P}$ gives
\begin{equation}
\langle {\cal P} \rangle_r = \left\langle \frac{1}{\Delta t} \int_{-\frac{\Delta t}{2}}^{\frac{\Delta t}{2}}
    \,dt\, \left(I(t)-I_0\right)^2 \right\rangle_r \simeq S_I(\omega) \Delta f
\label{av-power}  .
\end{equation}
The second moment,  $\langle {\cal P}^2 \rangle_r$,  can be evaluated
for a Gaussian random process $I(t)$, using stationarity:
\begin{eqnarray}
&&\langle {\cal P}^2 \rangle_r = \left\langle \left[
\frac{1}{\Delta t} \int_{-\frac{\Delta t}{2}}^{\frac{\Delta t}{2}}
    \,dt\, \left(I(t)-I_0\right)^2 \right]^2\right\rangle_r
\nonumber\\
&& \simeq  \frac{1}{\Delta t}\, \int_{-\infty}^{\infty} d\tau K_{I^2}(\tau) + \left(S_I(\omega) \Delta f\right)^2
\label{av-power2}  ,
\end{eqnarray}
where $K_{I^2}(\tau)$ is the correlation function of the ``power signal''
${\tilde I}^2(t) - \langle {\tilde I}^2(t) \rangle_r$
\begin{equation}
K_{I^2}(\tau) \equiv
\left\langle \left[{\tilde I}^2(t) - \langle {\tilde I}^2 \rangle_r \right]\,
\left[{\tilde I}^2(t+\tau) - \langle {\tilde I}^2 \rangle_r \right] \right\rangle_r
\label{cor-I2}  .
\end{equation}
The variance of the fluctuating power, Eq.\,(\ref{av-power2}), is expressed through
the ``second'' spectral density
$S_{I^2}(\omega) \equiv 2 \int_{-\infty}^{\infty} d\tau \, K_{I^2}(\tau) e^{\imat \omega \tau}$,
which for a Gaussian process $I(t)$ is related to the ordinary one.\cite{Kogan}
In particular, $S_{I^2}(0) = 2 [S_I(\omega)]^2 \Delta f$ and we obtain
\begin{equation}
\Delta {\cal P}^2 \equiv \langle {\cal P}^2 \rangle_r - \langle {\cal P} \rangle^2_r
 = \frac{1}{\Delta t} \, [S_I(\omega)]^2 \, \Delta f
\label{var-power}  .
\end{equation}

For a more general frequency filter $\alpha(f)$ the power is distributed
with average and variance respectively given as
$\langle {\cal P} \rangle_r = \int_{0}^{\infty} \alpha(f) S_I(\omega) df$ and
$\Delta {\cal P}^2 = \frac{1}{\Delta t} \int_{0}^{\infty} \alpha^2(f) \, [S_I(\omega)]^2 \, df$
that generalize Eqs.\,(\ref{av-power}),(\ref{var-power}).

    We are interested in measuring $I(t)$ to distinguish between two hypotheses:
(1) flat power spectrum $S_I(\omega)=S_0$ and
(2) power spectrum with a peak $S_I(\omega)=S_0 + S_{peak}(f)$.
Assuming  Gaussian distribution of the power signal we can
estimate the measurement time $\tau_{meas}$ at
which the variances $\Delta {\cal P}_1^2[\alpha(f)]$, $\Delta {\cal P}_2^2[\alpha(f)]$,
become small enough to distinguish between the corresponding averages
$\langle {\cal P}_1 \rangle_r$, $\langle {\cal P}_2 \rangle_r$.
For a ``signal-to-noise'' ratio of $1$, $\tau_{meas}$ is expressed as (see also Ref.\onlinecite{Kor-99-01})
\begin{eqnarray}
&& \tau_{meas} = \frac{ \left[ \,\sqrt{\Delta {\cal P}_1^2[\alpha(f)] }
         + \sqrt{\Delta {\cal P}_2^2[\alpha(f)] } \, \right]^2 }
{ \left[\int_{0}^{\infty} \alpha(f) S_{peak}(f) \, df \right]^2 }
\label{tau-m}  .
\end{eqnarray}
Optimization over the filter requires $\delta \tau_{meas} /\delta \alpha(f) =0$,
which leads to the filter form
\begin{equation}
\alpha(f) = \frac{S_{peak}(f)}{1 + a (1 + S_{peak}(f)/S_0)^2}
\label{opt-filter}
\end{equation}
that is close to the peak form itself while further optimization on
the parameter $a$ is needed.  However, for a small peak height,
$h \approx S_{peak}(\Omega)/S_0 \ll 1$, the result is independent
of $a$ and reads
\begin{equation}
\tau_{meas} = \frac{32}{\Gamma h^2}.
\label{tau-m-small-eta}
\end{equation}
Note that small height translates to small detector ideality
(for one qubit $h\leq 4 \eta$ while for two qubits $h\leq 32 \eta/3$).
To get an exact result for large or small height, one employs integration
by residues and find a cumbersome expression that can only be managed numerically.
Numerically, the calculation suggests that $a_{opt}$ is of the order of $1$.
An approximation
with accuracy better than $30\%$ even at large $\eta$ is given by
\begin{equation}
\tau_{meas} = \frac{32}{\Gamma h^2}\,\left(1+\frac{3 h}{4} + \frac{7 h^2}{64} \right)
\label{tau-m-approx}
\end{equation}
which means that $\tau_{meas}$ may become an order of magnitude larger
than the simple estimate, Eq.\,(\ref{tau-m-small-eta}).


\begin{thebibliography}{99}



\bibitem{BraginskyKhalili}
        V.~B. Braginsky and F.~Ya. Khalili,
        {\it Quantum measurement}
        (Cambridge Univ. Press, Cambridge, UK, 1992).



\bibitem{Joos} 
E. Joos, H.D. Zeh, C. Kiefer, D. Giulini, J. Kupsch, and I.-O. Stamatescu,
  {\it Decoherence and appearance of classical world in quantum theory},
  (Springer, Heidelberg, 2003).



\bibitem{Nakazato} 
M. Namiki, S. Pascazio, and H. Nakazato,
  {\it Decoherence and Quantum measurements},
  (World Scientific, Singapore, 1997).



\bibitem{Khalfin} L.S. Khalfin, JETP Lett. {\bf 8}, 63 (1968).


\bibitem{MisraSudarshan} B. Misra and E. C. G. Sudarshan, J. Math. Phys. {\bf 18}, 756 (1977);
C.B Chiu, E. C. G. Sudarshan, and B. Misra, Phys. Rev. D {\bf 16}, 520 (1977).


\bibitem{Peres} A. Peres, Am. J. Phys. {\bf 48}, 931 (1980).


\bibitem{Joos1} E. Joos, Phys. Rev. D {\bf 29}, 1626 (1984).



\bibitem{QZE-development} R. J. Cook, Phys. Scr. T {\bf 21}, 49 (1988);
    G. J. Milburn, J. Opt. Soc. Am. B {\bf 5}, 1317 (1988);
    A. Beige and G. C. Hegerfeldt, J. Phys. A {\bf 40}, 1323 (1997);
    L. S. Schulman, Phys. Rev. A {\bf 57}, 1509 (1998);
    A. G. Kofman and G. Kurizki, Nature {\bf 405}, 546 (2000);
    P. Facchi, H. Nakazato, S. Pascazio,
               Phys. Rev. Lett. {\bf 86}, 2699 (2001).


\bibitem{another} An alternative mechanism for stabilization of a quantum state
    is provided by a strong coupling to a noisy environment\cite{noise}.
    Suppression of the quantum transition in a two-level macroscopic quantum system
    due to large dissipation is discussed, e.g., in Ref.\onlinecite{Leggett-DirCondMatt}.


\bibitem{noise} M. Simonius, Phys. Rev. Lett. {\bf 40}, 980 (1978);
    R.A. Harris and L. Stodolsky, Phys. Lett. B {\bf 116}, 464 (1982);
    Ph. Blanchard, G. Bolz, M. Cini, G.F. De Angelis, and M. Serva,
    J. of Stat. Phys., {\bf 75}, 749 (1994).


\bibitem{Leggett-DirCondMatt} A.~J.~Leggett in {\it Directions in Condensed Matter Physics},
                     eds. G. Grinstein and G. Mazenko, (World Scientific, Singapore, 1986).


\bibitem{Itano} W. M. Itano, D. J. Heinzen, J. J. Bollinger, and D. J. Wineland,
                Phys. Rev. A {\bf 41}, 2295 (1990).
\bibitem{Fischer} M. C. Fischer, B. Guti\'{e}rrez-Medina, and M.G. Raizen,
                  Phys. Rev. Lett. {\bf 87}, 040402 (2001).


\bibitem{superqubits}  Y. Nakamura, Yu. A. Pashkin, and J. S. Tsai,
        Nature (London) {\bf 398}, 786 (1999);
        C. H. van der Wal, A.C.J. ter Haar, F. K. Wilhelm, R. N. Schouten,
        C.J.P.M. Harmans, T. P. Orlando, S. Lloyd, and J. E. Mooij,
        Science {\bf 290}, 773 (2000);
        J. R. Friedman, V. Patel, W. Chen, S. K. Tolpygo,
        and J. E. Lukens, Nature (London) {\bf 406}, 43 (2000);
    D. Vion, A. Aassime, A. Cottet, P. Joyez, H. Pothier,
    C. Urbina, D. Esteve, and M. H. Devoret, Science {\bf 296}, 886 (2002);
    J. M. Martinis, S. Nam, J. Aumentado, and C. Urbina,
        Phys.\ Rev.\ Lett. {\bf 89}, 117901 (2002).


\bibitem{QDots}  E.~Buks, R.~Schuster, M.~Heiblum, D.~Mahalu, and V.~Umansky,
             Nature {\bf 391}, 871 (1998);
        T. Hayashi, T. Fujisawa, H. D. Cheong, Y. H. Jeong, and
        Y. Hirayama,  Phys.\ Rev.\ Lett. {\bf 91}, 226804 (2003);
        W. Lu, Z. Ji, L. Pfeifer, K. W. West, and A. J. Rimberg,
        Nature {\bf 423}, 422 (2003);
        N. J. Graig, J. M. Taylor, E. A. Lester, C. M. Marcus, M. P. Hanson, and A. C. Gossard,
        Science {\bf 304}, 565 (2004).


\bibitem{Averin-SolSt}
        D. V. Averin, Fortschr. Phys. {\bf 48}, 1055 (2000).



\bibitem{SET} M. H. Devoret and R.~Schoelkopf, Nature {\bf 406}, 1039 (2000);
            T.~Duty, D.~Gunnarsson, K.~Bladh, and P.~Delsing,
            Phys.\ Rev.\ B {\bf 69}, 140503(R) (2004);
            A.~Guillaume, J.~F.~Schneiderman, P.~Delsing, H.~M.~Bozler,
            and P.~M.~Echternach, Phys.\ Rev.\ B {\bf 69}, 132504 (2004).


\bibitem{Ilichev1} E. Il'ichev, N. Oukhanski, A. Izmalkov, Th. Wagner, M. Grajcar, H.-G. Meyer, A.Yu. Smirnov,
             Alec Maassen van den Brink, M.H.S. Amin, and A.M. Zagoskin, Phys.\ Rev.\ Lett. {\bf 91}, 097906 (2003).


\bibitem{vonNeumann} J. von Neumann, {\it Mathematical Foundations of
        Quantum Mechanics} (Princeton Univ. Press, Princeton, 1955).



\bibitem{Mensky} M. B. Mensky, Phys.\ Rev.\ D {\bf 20}, 384 (1979).


\bibitem{Gisin} N. Gisin, Phys.\ Rev.\ Lett. {\bf 52}, 1657 (1984).


\bibitem{Caves-86} C. M. Caves, Phys.\ Rev.\ D {\bf 33}, 1643 (1986).



\bibitem{Carmichael} H. J. Carmichael, {\it An open system approach
        to quantum optics}, Lecture notes in physics (Springer, Berlin,
        1993).


\bibitem{Plenio} M. B. Plenio and P. L. Knight, Rev.\ Mod.\ Phys. {\bf 70},
    101 (1998).


\bibitem{Wiseman} H. M. Wiseman and G. J. Milburn, Phys. Rev. Lett.
        {\bf 70}, 548 (1993); Phys. Rev. A {\bf 49}, 1350 (1994).



\bibitem{Kor-99-01} A. N. Korotkov, Phys.\ Rev.\ B {\bf 60}, 5737 (1999);
        Phys. Rev. B {\bf 63}, 115403 (2001).



\bibitem{Kor-rev} A. N. Korotkov, cond-mat/0209629, in {\it Quantum noise}
    in mesoscopic physics, edited by Yu. V. Nazarov (Kluwer,
    Netherlands, 2003), p. 205.



\bibitem{Goan} H.-S. Goan, G. J. Milburn, H. M. Wiseman, and H. B. Sun,
        Phys.\ Rev.\ B {\bf 63}, 125326 (2001).


\bibitem{Goan-2} H.-S. Goan and G. J. Milburn, Phys.\ Rev.\ B {\bf 64}, 235307 (2001);
                 N. P. Oxtoby, H. B. Sun, and H. M. Wiseman,
                 J. Phys.: Condens. Matter {\bf 15}, 8055 (2003).


\bibitem{Kor-osc} A. N. Korotkov and D. V. Averin, Phys.\ Rev. B {\bf 64}, 165310 (2001);
                  A. N. Korotkov, Phys. Rev. B {\bf 63}, 085312 (2001).


\bibitem{Makhlin} Yu. Makhlin, G. Sch\"{o}n, A. Shnirman,
        Phys. Rev. Lett. {\bf 85}, 4578 (2000).


\bibitem{ShepelyanskyDetect} J. W. Lee, D. V. Averin, G. Benenti, and D. Shepelyansky,
                            Phys.\ Rev. A {\bf 72}, 012310 (2005).  


\bibitem{RKMineq} R. Ruskov, A. N. Korotkov, and A. Mizel, quant-ph/0505094.



\bibitem{RusKor-ent} R.~Ruskov and A.~N.~Korotkov,
                   Phys.\ Rev. B {\bf 67}, 241305(R) (2003).



\bibitem{prl-quadr} W.~Mao, D.V. Averin, R.~Ruskov, and A.~N.~Korotkov,
                   Phys.\ Rev.\ Lett. {\bf 93}, 056803 (2004).


\bibitem{Mao} W.~Mao, D.~V.~Averin,  F.~Plastina, and R.~Fazio,
                               Phys. Rev. B {\bf 71}, 085320 (2005).


\bibitem{Gurvitz} S. A. Gurvitz, Phys. Rev. B {\bf 56}, 15215 (1997).


\bibitem{AverinFazio} D.~V.~Averin and R.~Fazio, JETP Letter, {\bf 78}, 1162 (2003).



\bibitem{Kor-ent} A. N. Korotkov, Phys. Rev. A {\bf 65}, 052304 (2002).


\bibitem{Kor-nonideal} A. N. Korotkov, Phys. Rev. B {\bf 67}, 235408, (2003).


\bibitem{Ruskov-osc} R. Ruskov, K. Schwab, and A. N. Korotkov,
                Phys. Rev. B {\bf 71}, 235407, (2005).


\bibitem{AverinSukhorukov} D.~V.~Averin and E. V. Sukhorukov,
            Phys. Rev. Lett. {\bf 95}, 126803 (2005).

\bibitem{Oksendal} B.{\O}ksendal, {\it Stochastic differential equations}
    (Springer, Berlin, 1998).

\bibitem{GardinerZoller} C. W. Gardiner and P. Zoller, {\it Quantum noise}
    (Springer, Berlin, 2000).

\bibitem{Leggett} A.~O.~Caldeira and A.~J.~Leggett,  {\em Ann. Phys.} (N.Y.) {\bf 149},
        374 (1983);
        W.~H.~Zurek,  {\em Physics Today} {\bf 44}, 36 (1991).


\bibitem{Pilgram} S. Pilgram and M. B\"{u}ttiker, Phys. Rev. Lett. {\bf 89}, 200401 (2002).


\bibitem{Clerk} A. A. Clerk, S. M. Girvin, and A. D. Stone, Phys. Rev. B {\bf 67}, 165324 (2003).



\bibitem{RusKor-sp} R.~Ruskov and A.~N.~Korotkov, Phys. Rev. B {\bf 67}, 075303 (2003).



\bibitem{ideality}
Experimentally\cite{QDots} QPC  is close to an ideal
detector. An SET is usually highly non-ideal\cite{Kor-99-01},
however it may reach ideality close to 1 in the co-tunneling or Cooper pair tunneling regime\cite{cotun}.


\bibitem{cotun} A. B. Zorin, Phys. Rev. Lett. {\bf 76}, 4408 (1996);
                D. V. Averin, cond-mat/0010052 (unpublished);
             A. A. Clerk, S. M. Girvin, A. K. Nguyen, and A. D. Stone, Phys. Rev. Lett. {\bf 89}, 176804 (2002).


\bibitem{Mabuchi}  H. Mabuchi, H. M. Wiseman,
                  Phys. Rev. Lett. {\bf 81}, 4620 (1998); Erratum-ibid. {\bf 82}, 1798 (1999).


\bibitem{objectivity} We also checked by numerical simulations of the measurement process
that two observers, who start with different initial knowledge about
the system (i.e., different initial
density matrices $\rho_{ij}(0)$) will agree with each other about
the state evolution after such measurement time
(compare with Ref.\onlinecite{Ruskov-osc}).



\bibitem{Bulaevskii} L. N. Bulaevskii and G. Ortiz,
                  Phys.\ Rev.\ Lett. {\bf 90}, 040401 (2003).


\bibitem{Shnirman-sp} A. Shnirman, D. Mozyrsky, and I. Martin,
                   Europhys. Lett., {\bf 67} (5), 840 (2004).


\bibitem{Davies} E. B. Davies, {\it Quantum Theory of Open Systems},
               (Academic, London, 1976).

\bibitem{Gardiner} C. W. Gardiner, {\it Handbook of Stochastic methods}
    (Springer, Berlin, 1983).

\bibitem{Kogan} Sh. Kogan, {\it Electronic noise and fluctuations in solids}
    (Cambridge University Press, Cambridge, USA, 1996).

\bibitem{Machlup} S. Machlup, J. Appl. Phys. {\bf 25}, 341 (1954).


\end{thebibliography}
\end{document}